\documentclass[12pt,twoside]{JHEP3}

\usepackage{cite}
\usepackage{amsmath}
\usepackage{amssymb}
\usepackage{mathtools}
\usepackage{mathrsfs}
\usepackage{bm}
\usepackage{graphicx}

\newcommand{\ubar}{\bar{u}}

\newcommand{\beq}{\begin{equation}}
\newcommand{\eeq}{\end{equation}}
\newcommand{\beqa}{\begin{eqnarray}}
\newcommand{\eeqa}{\end{eqnarray}}
\newcommand{\nonr}{\nonumber}

\title{An effective gauge-Higgs operators analysis of new physics associated with the Higgs}
\author{We-Fu Chang, Wei-Ping Pan, and Fanrong Xu\\
	Department of Physics, National Tsing Hua University, HsinChu 300, Taiwan\\
	E-mail: \email{wfchang@phys.nthu.edu.tw}, \email{two.joker@gmail.com}, \email{frxu@phys.nthu.edu.tw}}

\abstract
{We study the new physics(NP) related to the recent discovered $125$ GeV Higgs by employing an important subset of
the standard model(SM) gauge invariant dimension-six operators constructed by the the SM Higgs and  gauge fields.
Explicitly, we perform a model-independent study on the production and decays of the Higgs, the electric dipole moments(EDMs) of the neutron and the electron, and we take into account the anomalous magnetic dipole moments of muon  and electron as well.
 We find that,  even all Higgs decay channels agree
with the SM predictions, the SM theoretical uncertainties provide a lot of room to host NP associated with the 125 GeV boson.
A linear relation is revealed in our numerical study that  $\mu_{ZZ}\simeq \mu_{WW}$ and $ 0.6 \lesssim \mu_{ZZ,WW} \lesssim 1.4$ at $95\%$CL with or without the EDMs constraints.
The neutron and electron EDMs  severely constrain the relevant Wilson coefficients. Therefore the CP violating components in the $h\rightarrow WW, ZZ$ channels  are too small, $\sim{ \cal O}(10^{-5})$,  to be detected at the LHC.
However, we point out that even the parity of the 125GeV boson has been largely determined to be even in the $h\to ZZ$ channel, one should pay special attention to the potentially large CP violation in the $h\to \gamma\gamma$ and $h\to \gamma Z$ channels.
This should be seriously checked in the future spin correlation experiments.
}


\begin{document}


\section{Introduction}

The high energy physics community has been excited about the recent finding of a standard model(SM) like Higgs boson $h$, the  final piece of the SM we had long craved for,  at the mass around 125 GeV at the LHC \cite{Higgs-ATLAS, Higgs-CMS }.
So far,  except the Higgs diphoton decay which is $\sim 1\sigma$ higher than its SM prediction,
all other Higgs decay modes agree with the SM predictions within the experimental accuracies\cite{CMS-bosonic,CMS-bosonic-QCD,CMS-fermionic,CMS-gamma,CMS-ZZ,CMS-WW,CMS-tautau,CMS-gammaZ,ATLAS-bosonic,ATLAS-bosonic-QCD,ATLAS-fermionic,LHC-fermionic,ATLAS-gammaZ,ATLAS-gamma,ATLAS-ZZ,ATLAS-WW,ATLAS-tautau,ATLAS-bb}.
The diphoton excess could  be just the statistical fluctuation or due to some unknown systematic error, or
it indicates  the existence of new physics(NP) especially those connecting the Higgs and gauge sectors.
 Since last year, there have been many discussion and speculation about the diphoton excess, see for example\cite{many}.
Whether the diphoton excess is due to new physics or not will be settled down by more experimental efforts in the coming years.
Although the statistics is still weak, the current Higgs decay data has already set some limits to the NP associated with the
Higgs boson.
On the other hand, we should also bear in mind that  the uncertainty of the SM theoretical prediction for the Higgs productions
range from $\sim 2\%$ for the vector boson fusion(VBF) to $\sim 15\%$ for the gluon-gluon fusion(GF)\cite{Higgs-Xsec}, the dominate Higgs production mechanism at the LHC, see Table \ref{tb:SM-xsec}.  As for the decay branching ratios, the theoretical uncertainties range
 from $\sim 3\%$ for $h\rightarrow b{b}$ to $\sim 12\%$ for $h\rightarrow c\bar{c}$ \cite{Higgs-Xsec}, see Table \ref{tb:SM-br}.
So even the experimental precision can be improved to reach a few percent level in the future, still, we will not be able to conclude the total absence of NP associated with the $\sim 125$ GeV boson.  The interesting question will then be, `` How much room for new physics is still allowed due to the intrinsic SM theoretical uncertainty?''  The main purpose of this paper is aiming for a model-independent  constraint on the new physics associated with the Higgs  based on the current and future data.

Meanwhile, the mass scale of exotic degrees of freedom(DOF) have been pushed to be greater than around $(0.4-8)$ TeV
in various scenarios with different assumptions \cite{exoticATLAS, exoticCMS}. Although   a general analysis is lacking, it is now wildly believed that the cutoff scale, $\Lambda$, for  physics beyond SM  begins to show up, should be much greater than  the electroweak symmetry breaking
scale, $v\sim 250$ GeV.
Motivated by the diphoton excess and the exotic DOF(s) which is(are) heavier than the electroweak scale, it is reasonable to assume that the new physics effects can be captured and described by the dimension-six (dim-6) gauge-Higgs operators.
We assume that the NP decouples at $\Lambda\gg v$ where the SM electroweak symmetry is not yet broken.  Below  the cutoff,  there are only SM DOFs and the SM gauge symmetry are still honored by the low energy effective theory. The information and effects of NP
are encoded in the form of the effective operators and their corresponding Wilson coefficients; see \cite{BW86, dim6-10} for the early general discussion.
It is not new to study the Higgs physics by using the gauge-Higgs operators,
see  \cite{ZY,Pospelov, MWPLB, RGE-Manohar,anom-coupling} for the previous studies along the same line
 and \cite{similar-eft} for recent update including the Moriond 2013 data. However, some constraints have been over looked by those
 authors.  The current analysis is so far the most comprehensive one to our best knowledge. We found those contributions which had been over looked indeed play a significant role to constrain NP.

In this work we should perform a model-independent analysis based on a subset of
SM gauge invariant dim-6 gauge-Higgs operators. The presence of these dim-6 operators modified
the gauge-Higgs couplings so as the Higgs decay and production at the tree-level.
We should discuss the constraints on the relevant Wilson coefficients based on the LHC
Higgs  data  and compare it to a benchmark case where all Higgs decay channels sit at the predicted  SM values and the SM theoretical uncertainties are used as the experimental errors.  We should show that indeed the parasitic NP can live comfortably within the
SM theoretical uncertainties without much upsetting the global fit.

In addition to the tree-level processes, the coefficients of CP odd operators can be further constrained by
 estimating the loop-induced EDM at low energy. The apparent drawback of using effective operators is that we do not have a sensible prediction of the quantum effects at the loop level. The dim-6 gauge-Higgs operators do not close under the RG running\cite{RGE-Manohar}.
 To proceed, we  consider the class of new physics in which
the fermion EDM operators can be ignored at $\Lambda$\footnote{  This working assumption is also reasonable from phenomenology point of view.
    The current experimental limits require that the cutoff scale of these EDM operators
    to be greater than $10^4$ TeV if the corresponding Wilson coefficients are all $\sim {\cal O}(1)$. Here we further assume the relevant Wilson coefficients
for fermion EDMs generated by higher loops at $\Lambda$ or RG running are negligible.} and the divergent part of loop integral be removed by the counter terms. Practically, we use dimensional regularization to calculate the leading contribution  of CP-odd operator
to the fermion EDM in the modified minimal subtraction ($\overline{MS}$) scheme.
Similarly, just for a ballpark estimation  of how the $(g-2)$ will limit the NP,  without any better argument, we also assume  the charged lepton $(g-2)$ is negligible at $\Lambda$.  We calculate the loop induced muon and electron $(g-2)$ by using the gauge-Higgs operators at low energy and further constrain the relevant CP even Wilson coefficients.
At the end, we will discuss a UV-complete toy model as an example for this assumption to work.
But we should keep in mind that this assumption is  not valid for the general case.

The paper is organized as follows.
 In section 2,  we will lay out the important subset of the dim-6 gauge-Higgs operators.
The needed Feynman rules for later calculation will be summarized there.
 The  modifications to the Higgs decay and production due to these  gauge-Higgs operators at the tree-level will be discussed in Section 3.
 The 1-loop contributions to EDM and g-2 will be given in Section 4.
 In Section 5, we present the numerical results and some remarks from the global fitting.
In section 6, we briefly discuss how  a degenerate solution can be resolved by the Higgs pair production cross section.
As an example of the UV origin of the gauge-Higgs operators
 and to illustrate in what condition our analysis is valid, two simple  models with scalar color octet, but in different $SU(2)$ representations, will be discussed.
 A brief summary will be given there as well.
Finally,  some technical details will be collected in Appendix.

\section{Effective Lagrangian}
As discussed in previous section, we assume that the NP associated with the SM  Higgs production and decay could be largely
captured  and described by a set of   $SU(3)_C\times SU(2)_L\times U(1)_Y$ invariant dim-6  gauge-Higgs effective  operators
and their Wilson coefficients\cite{MW, Pospelov}
\begin{equation}
\mathscr{L}_{NP} = \sum_i (c_i\mathcal{O}_i +\tilde{c}_i\tilde{\mathcal{O}}_i)\;\;\;(\mbox{i=1, 2, 12, 3} )\, .\label{Lan}
\end{equation}
The relevant effective gauge-Higgs operators are
\begin{align}
&\mathcal{O}_1=\frac{g_1^2}{2\Lambda^2} H^\dagger H B_{\mu\nu}B^{\mu\nu},~~
\tilde{\mathcal{O}}_1=\frac{g_1^2}{2\Lambda^2} H^\dagger H B_{\mu\nu}\widetilde{B}^{\mu\nu}\,,
\nonumber\\
&\mathcal{O}_2=\frac{g_2^2}{2\Lambda^2} H^{\dagger} H W^a_{\mu\nu} W^{a\,\mu\nu},~~
\tilde{\mathcal{O}}_2= \frac{g_2^2}{2\Lambda^2}H^{\dagger} H W^a_{\mu\nu} \widetilde{W}^{a\,\mu\nu}\,,
\nonumber\\
&\mathcal{O}_{12}= \frac{g_1 g_2}{2\Lambda^2}H^\dagger \tau^a H B_{\mu\nu} {W}^{a\,\mu\nu},~~
\tilde{\mathcal{O}}_{12}= \frac{g_1 g_2}{2\Lambda^2}H^\dagger \tau^a H B_{\mu\nu} \widetilde{W}^{a\,\mu\nu}\,,
\nonumber\\
&\mathcal{O}_3= \frac{g_3^2}{2\Lambda^2}H^\dagger H G_{\mu\nu}^A G^{A\,\mu\nu},~~
\tilde{\mathcal{O}}_3=\frac{g_3^2}{2\Lambda^2} H^{\dagger} H G^A_{\mu\nu} \widetilde{G}^{A\,\mu\nu}\,,
\label{eq:GHOperator}
\end{align}
where the dual field strength tensor is defined as $\widetilde{F}_{\mu\nu}=\frac12\epsilon_{\mu\nu\lambda\sigma}F^{\lambda\sigma}$, $(F=B, W, G)$,
and the others are in the standard notations. Note that we have absorbed the SM gauge couplings and the cutoff $\Lambda$ into the definition of
the operators.
The operator ${\cal O}_{12}$ gives the direct  $SU(2)$-$U(1)_Y$ gauge mixing and modifies the oblique parameter $S$\cite{Oblique} at the tree level\cite{MW}
 \begin{equation}
\triangle S=\frac{8\pi v^2}{\Lambda^2}c_{12}\,,
\end{equation}
 when the two Higgs fields are replaced by their VEVs.
The current $S$ parameter bound, $S=0.00^{+0.11}_{-0.10}$\cite{PDG2012}, already sets a strong limit on $c_{12}$.
Similarly, the $\tilde{\mathcal{O}}_3$ operator yields an effective QCD $\theta_{QCD}$-term  and thus $\tilde{c}_3$ is strongly
constrained by the neutron EDM. Therefore, we will set $c_{12}=\tilde{c}_3=0$ in our  global analysis which
will be discussed in the numerical section.

There are three more dim-6 gauge-Higgs operators which give rise to the tree-level modification to the coupling between
Higgs and gauge fields:
\begin{equation}
\label{eq:minoOP}
(D_\mu H)^\dag (D_\nu H) B^{\mu\nu}\,,\;
(D^\mu H)^\dag \sigma^a (D^\nu H) W^{a}_{\mu\nu}\,,\;
| H^\dag D_\nu H |^2\, .
\end{equation}
However, these operators are severely constrained by electroweak precision tests;
for example, the SM gauge boson mass matrix will be altered  by these operators at tree-level.
Since we are focusing on the Higgs physics at the LHC and it is safe to ignore these operators,  we  confine our analysis with six out of the eight gauge-Higgs operators given in Eq.(\ref{eq:GHOperator}).

\begin{table}[htb]
\renewcommand{\arraystretch}{1.2}\addtolength{\arraycolsep}{3pt}
\caption[]{Feynman rules for the $hV^\mu(k_1)V'^\nu(k_2)$ vertices,  where $( k_1, k_2 )$ are entering the vertices.
 }
$$
\begin{array}{c|c|c}
\hline
 &S^{\mu\nu}(k_1,k_2) & P^{\mu\nu}(k_1,k_2) \\
\hline\hline
 h\gamma\gamma &  i\, \frac{2v g_2^2s_W^2}{\Lambda^2}a_1 &  i\, \frac{2vg_2^2 s_W^2}{\Lambda^2}  \tilde{a}_1       \\
 h\gamma Z     &  i\,\frac{vg_2^2}{\Lambda^2}a_2      & i\,\frac{vg^2_2}{\Lambda^2} \tilde{a}_2      \\
 h gg  &     i\,\frac{2vg_3^2}{\Lambda^2} c_3     &   i\, \frac{2vg_3^2}{\Lambda^2} \tilde{c}_3\simeq 0      \\
  h ZZ  &   i\,\frac{2vg_2^2}{\Lambda^2} a_4   &   i\,\frac{2vg_2^2}{\Lambda^2} \tilde{a}_4    \\
  hWW  &        i\, \frac{2vg_2^2}{\Lambda^2}c_2      &     i\,\frac{2vg_2^2}{\Lambda^2}\tilde{c}_2              \\
 \hline
\end{array}
$$
\label{interaction-hVV}
\end{table}
The relevant $hV^\mu(k_1)V'^\nu(k_2)$ Feynman rules  are summarized in Table \ref{interaction-hVV}
where $( k_1, k_2 )$ are the 4-momentum carried by the gauge bosons, and $(\mu, \nu)$ are the corresponding Lorentz indices.
The two gauge invariant form factors are defined as
 \begin{equation}
 S^{\mu\nu}(k_1,k_2) = k_2^\mu k_1^\nu - k_1\cdot k_2g^{\mu\nu}\,,\;
P^{\mu\nu}(k_1,k_2)=\epsilon^{\alpha\beta\mu\nu} k_{1\alpha} k_{2\beta}\,.
 \end{equation}
Also we define
 \begin{align}
 \label{eq:def_coef_a}
 &a_1\equiv c_1+c_2-c_{12}\simeq  c_1+c_2 \,, \; \tilde{a}_1\equiv \tilde{c}_1+ \tilde{c}_2 -\tilde{c}_{12}\nonumber\,,\\
&a_2\equiv s_{2W}\left(c_2 -c_1 t_W^2
                   -\frac12 c_{12}(1-t_W^2) \right)\simeq  s_{2W}\left(c_2 -c_1 t_W^2
                    \right) \,, \nonumber\\
&\tilde{a}_2\equiv s_{2W} \left(\tilde{c}_2 -\tilde{c}_1 t_W^2
-\frac12 \tilde{c}_{12}(1-t_W^2) \right)\nonumber\,,\\
&a_4\equiv c_W^2 (c_2 +c_1 t_W^4 +  c_{12} t_W^2 )\simeq c_W^2 (c_2 +c_1 t_W^4  )\,,\nonumber\\
&\tilde{a}_4\equiv c_W^2 (\tilde{c}_2+ \tilde{c}_1 t_W^4
+\tilde{c}_{12} t_W^2)\,,
\end{align}
where   $\sin\theta_W$ is denoted as $s_W$ for notational convenience.  Similarly, we adopt the following abbreviations:  $\cos\theta_W\rightarrow c_W$, $\sin 2\theta_W \rightarrow  s_{2W}$,  $\tan\theta_W\rightarrow t_W$ etc.
From Table \ref{interaction-hVV}, one can easily obtain the corresponding $hhVV'$ Feynman rules by replacing the VEV, $v$, by the Higgs field.
Due to the non-Abelian nature of $SU(2)_L$, the operators ${\cal O}_{2}, \tilde{{\cal O}}_{2,12}$  also give rise to the tree-level
$\gamma W W$ coupling when both SM Higgs fields take their VEVs.
The extra $\gamma WW$ interaction has been overlooked in the previous studies.
However it gives nonzero contributions to the fermion EDM and $(g-2)$.
 Explicitly, after electroweak SSB, the relevant Lagrangian is
 \begin{equation}
\delta\mathscr{L}_{\gamma VV} =~~~ c_2 g_2^2 \frac{v^2}{4\Lambda^2} W^{a}_{\mu\nu} W^{a,\mu\nu }
+ \tilde{c}_2 g_2^2 \frac{v^2}{4\Lambda^2} W^{a}_{\mu\nu} \widetilde{W}^{a,\mu\nu } 
- \tilde{c}_{12} g_1 g_2 \frac{v^2}{4\Lambda^2} B_{\mu\nu} \widetilde{W}^{3,\mu\nu }\, .
\end{equation}
The $\tilde{c}_2$ term is equivalent to a total derivative which has no effect in the local perturbation calculation.
The  $c_2$ term modifies the canonical normalization of the  kinematic term of  $SU(2)$ gauge fields.
Thus the  $c_2$-corresponding $\gamma W W$ form factor is same as in the SM but the coupling is now the SM one times $-c_2 g_2^2 v^2/\Lambda^2 $.

For the triple $\gamma^\alpha(k_1) W^{+,\beta}(k_2) W^{-,\lambda}(k_3)$ coupling, where all three momenta are entering the vertex, the corresponding  Feynman rules  for   $\tilde{c}_{12}$ can be spelled out
\begin{equation}
 \tilde{\Gamma}_6^{\alpha\beta\lambda}= i\, \frac{g_2^3 s_W v^2}{2\Lambda^2}\tilde{c}_{12} \epsilon^{\mu\alpha\beta\lambda} k_{1\mu}\, .
\end{equation}

\section{Higgs Production and decay}

The SM Higgs production cross sections and its branching ratios have been calculated and maintained by ``LHC Higgs Cross Section Working Group'' \cite{Higgs-Xsec}. For the later convenience, we collect the SM results in  Table \ref{tb:SM-xsec} and \ref{tb:SM-br}.
Note that the  inherent theoretical uncertainties of the SM  prediction for both Higgs production and decays range from a few percents to about $\sim 20\%$.

\begin{table}[htb]
\renewcommand{\arraystretch}{1.2}\addtolength{\arraycolsep}{3pt}
\caption[]{Cross sections of the SM Higgs boson at $125 \mathrm{GeV}$\cite{Higgs-Xsec},
where WH and ZH stand for the associated production with W and Z respectively. The uncertainties are shown in percentages. }
\label{tb:SM-xsec}
$$
\begin{array}{|l|c|c|c|}
\hline
\multicolumn{4}{|c|}
{\sigma(\mathrm{pb})}
\tabularnewline
\hline\hline
 & \mathrm{LHC}~ 7\,\mathrm{TeV} ~(\%) &\mathrm{LHC}~ 8\,\mathrm{TeV}~(\%) &\mathrm{LHC}~ 14\,\mathrm{TeV} ~(\%)\\
 \hline
\mathrm{GF} & 15.32^{+14.7}_{-14.9}    &  19.52^{+14.7}_{-14.7}   & 49.85^{+19.6}_{-14.6}  \\
\mathrm{VBF} &1.205^{+2.7}_{-2.4}     &  1.578^{+2.8}_{-3.0}    &  4.180^{+2.8}_{-3.0}    \\
\mathrm{WH}  &0.5729^{+3.7}_{-4.3}     &  0.6966^{+3.7}_{-4.1}   &  1.504^{+4.1}_{-4.4}    \\
\mathrm{ZH}  &0.3158^{+4.9}_{-5.1}     & 0.3943^{+5.1}_{-5.0}    &  0.8830^{+6.4}_{-5.5}     \\
\hline
\end{array}
$$
\end{table}

\begin{table}[htb]
\renewcommand{\arraystretch}{1.2}\addtolength{\arraycolsep}{3pt}
\caption[]{Branching ratios of different decay channels for the SM Higgs boson at $125 \mathrm{GeV}$\cite{Higgs-Xsec}.
The uncertainties are shown in percentages.}
\label{tb:SM-br}
$$
\begin{array}{|c|c|c|c|c|c|c|c|}
\hline
\multicolumn{8}{|c|}
{10^2\times\mathcal{B}(h\to ij)}
\tabularnewline
\hline\hline
bb (\%)& cc (\%) &\tau\tau (\%) & \gamma\gamma (\%) & ZZ (\%) & WW (\%) & gg (\%)& \gamma Z (\%) \\
\hline
57.7^{+3.2}_{-3.3} & 2.91^{+12.2}_{-12.2} & 6.32^{+5.7}_{-5.7} & 0.228^{+5.0}_{-4.9} & 2.64^{+4.3}_{-4.2}&
21.5^{+4.3}_{-4.2}& 8.57^{+10.2}_{-10.0}& 0.154^{+9.0}_{-8.8}\\
\hline
\end{array}
$$
\end{table}

Both CP-even and CP-odd operators affect the Higgs production and decay at tree level,
and the two kinds of contributions do not mix.
It is useful to define the following  ratios to characterize different  production and decay  channels of the SM Higgs
\begin{equation}
\alpha_{ij}=\frac{\Gamma(h\to i j)}{\Gamma^{\mathrm{SM}}(h\to i j)}\,,\qquad
\gamma_{XY}=\frac{\sigma(XY\to h)}{\sigma^{\mathrm{SM}}(XY\to h)}\, ,
\end{equation}
where $(i,j)$ and $(X,Y)$ stand for the final states and the initial particles respectively.
At tree-level, the gauge-Higgs operators have no effects on the SM Yukawa couplings so that $\alpha_{ff}=1$.
Since the gluon-gluon fusion(GF) is the dominate contribution to the Higgs production at the LHC, for simplicity
we will ignore the change to the other production channels due to the presence of the gauge-Higgs operators.
Next, we discuss how  the gauge-Higgs operators modify $\gamma_{gg}$  and the relevant $\alpha_{ij}$.

\subsection{Gluon fusion}
The effective operators which describe short distance features will not change gluon PDF and kinematics, so we just need to
focus on partonic cross section, which is
\begin{equation}
\hat{\sigma}_{\mathrm{LO}}(gg\to h)=\sigma_0^hm_h^2\delta(\hat{s}-m_h^2)
=\frac{\pi^2}{8m_h}\Gamma(h\to gg)\delta(\hat{s}-m_h^2)\, .
\end{equation}
Then the ratio of cross section is
\begin{equation}
\gamma_{gg}\equiv
\frac{\hat\sigma(gg\to h)}{\hat\sigma_{\mathrm{SM}}(gg\to h)}=\frac{\Gamma(h\to gg)}{\Gamma_{\mathrm{SM}}(h\to gg)}\, .
\end{equation}

In the SM, Higgs decays to two gluons via heavy quark loop, and the decay width is well-known\cite{Higgs-rev}
\begin{equation}
\Gamma(h\to gg)=
\frac{G_F\alpha_s^2m_h^3}{64\sqrt{2}\pi^3}\left|
A_{\frac12}(\tau_t)\right|^2\,,
\end{equation}
where $\tau_t=\frac{m_h^2}{4m_t^2}$, and $A_{\frac12}$ is given by
\begin{align}
&A_{\frac12}(\tau)=2\left[\tau+(\tau-1)f(\tau)\right]\tau^{-2}\,,\nonumber\\
&f(\tau)=\left\{\begin{array}{ll}
\arcsin^2\sqrt{\tau} & \tau\leq 1\\
-\frac14\left[\ln\frac{1+\sqrt{1-\tau^{-1}}}{1-\sqrt{1-\tau^{-1}}}-i\pi\right]^2 & \tau>1
\end{array}
\right.\, .
\end{align}

Incorporate the CP-even and CP-odd effective operator contributions, the total decay width can be calculated to be
\begin{equation}
\Gamma(h\to g g)=\frac{G_F\alpha_s^2m_h^3}{64\sqrt{2}\pi^3}\left(\left|A_{\frac12}(\tau_t)
+\frac{16\sqrt{2}\pi^2}{G_F\Lambda^2}c_3\right|^2
+\left|  \frac{16\sqrt{2}\pi^2}{G_F\Lambda^2}\tilde{c}_3 \right|^2
\right)\,,
\end{equation}
which agrees with \cite{MWPLB}. But as had already stated, the $\tilde{c}_3$ is severely  constrained by neutron EDM,
only the CP-even operator ${\cal O}_3$ is relevant to the ratio of gluon-gluon fusion cross section
\begin{equation}
\gamma_{gg}=\left|1
+\frac{16\sqrt{2}\pi^2}{G_F\Lambda^2 A_{\frac12}(\tau_t)}c_3\right|^2\,.
\end{equation}

\subsection{Diphoton decay}
The $h \rightarrow2\gamma$ decay width relates to the amplitude
\begin{equation}
\Gamma=\frac{1}{32\pi\, m_h}|\mathcal  {M}^{\gamma\gamma}|^2\,,
\end{equation}
where
\begin{equation}
\mathcal{M}^{\gamma\gamma}=\mathcal{M}_{\mathrm{SM}}^{h\to \gamma\gamma}+\frac{2v g_2^2s_W^2}{\Lambda^2}a_1 S^{\alpha\beta}(k_1,k_2)\epsilon_{1\alpha}^{*}\epsilon_{2\beta}^{*}
+
\frac{2vg_2^2 s_W^2}{\Lambda^2}  \tilde{a}_1  P^{\alpha\beta}(k_1,k_2)\epsilon^*_{1\alpha}\epsilon^*_{2\beta}\,,
\end{equation}
with the SM contribution \cite{Higgs-rev,W-loop}
\begin{align}
&\mathcal{M}^{h\to\gamma\gamma}_{\mathrm{SM}}=
M_1
 S_{\mu\nu}(k_1,k_2)
\epsilon^{\mu*}_1\epsilon^{\nu*}_2\,,\nonumber\\
&M_1=\frac{\alpha}{2\pi}\left(\sqrt{2}G_F\right)^{\frac12}
\left(
A_1(\tau_W)+\frac43 A_{\frac12}(\tau_t)
\right)\,,
\end{align}
where $A_{\frac12}$ is from top quark loop as given in the GF section, and $A_{1}$ is from the W boson loop,
\begin{equation}
A_1(\tau)=-\left[2\tau^2+3\tau+3(2\tau-1)f(\tau)\right]\tau^{-2}\,,
\end{equation}
and parameter $\tau_i=\frac{m_h^2}{4m_i^2}$.
It is straightforward to calculate  the ratio of decay rates
\begin{eqnarray}
\alpha_{\gamma\gamma}&=&\frac{|\mathcal{M}( h\to\gamma\gamma)|^2}{|\mathcal{M}_{\mathrm{SM}}(h\to\gamma\gamma)|^2}\\
&=&\left|1+ \frac{8\sqrt{2}\pi^2 a_1}{G_F\Lambda^2[A_1(\tau_W)+\frac43 A_{\frac12}(\tau_t)]}\right|^2+ \left| \frac{8\sqrt{2}\pi^2 \tilde{a}_1 }{G_F\Lambda^2[A_1(\tau_W)+\frac43 A_{\frac12}(\tau_t)]}\right|^2\, . \nonumber
\end{eqnarray}

\subsection{$h\to \gamma Z$}
The decay rate for $h\to \gamma Z$ in the rest frame of Higgs boson is
\begin{equation}
\Gamma=\frac{1}{16\pi m_h}\left(1-\frac{m_Z^2}{m_h^2}\right)|\mathcal{M}^{\gamma Z}|^2\,.
\end{equation}
The amplitude including high dimensional operators' contributions is
\begin{equation}
\mathcal{M}^{\gamma Z}=\mathcal{M}_{\mathrm{SM}}^{(h\to \gamma Z)}+
\frac{vg_2^2}{\Lambda^2}a_2  S^{\mu\nu}\epsilon^*_{1\mu} \epsilon^*_{2\nu}
+\frac{vg^2_2}{\Lambda^2} \tilde{a}_2 P^{\mu\nu}\epsilon^*_{1\alpha}
\epsilon^*_{2\beta}\,,
\end{equation}
and the SM contribution \cite{Higgs-rev,Higgs-hunter,HAZ} is given as
\begin{align}
&\mathcal{M}_{\mathrm{SM}}^{(h\to\gamma Z)}=M_2 S^{\mu\nu} \epsilon^*_{1\mu} \epsilon^*_{2\nu}\,,
\nonumber\\
&M_2
=\frac{\alpha}{2\pi}\left(\sqrt{2}G_F\right)^{\frac12}(A_F+A_W)\,,
\nonumber\\
&A_F=-\frac{2\hat{v}_t}{c_Ws_W}A^{\gamma Z}_{\frac12}(\tau_t,\lambda_t)\,,\;
A_W=-\frac{1}{s_W}A^{\gamma Z}_1(\tau_W,\lambda_W)\,,
\end{align}
where $\tau_i=\frac{m_h^2}{4m_i^2}$, $\lambda_i=\frac{m_Z^2}{4m_i^2}$ (i=$t,W$), and fermion vector coupling $\hat{v}_f=2I_f^3-4Q_fs_W^2$.
The relevant functions are
\begin{align}
&A^{\gamma Z}_{\frac12}(\tau,\lambda)=[I_1(\tau,\lambda)-I_2(\tau,\lambda)]\,,\nonumber\\
&A^{\gamma Z}_1(\tau,\lambda)=c_W\left\{4\left(3-t_W^2\right)I_2(\tau,\lambda)
+\left[ (1+2 \tau )t_W^2-(5+ 2\tau )\right]
I_1(\tau,\lambda)\right\}\,,\nonumber\\
& I_1(\tau,\lambda)=\frac{1}{2(\lambda-\tau)}+\frac{1}{2(\lambda-\tau)^2}
\left[f(\tau)-f(\lambda)\right]+\frac{\lambda}{(\lambda-\tau)^2}
\left[g(\tau)-g(\lambda)\right]\,,\nonumber\\
&I_2(\tau,\lambda)=-\frac{1}{2(\lambda-\tau)}\left[f(\tau)-f(\lambda)\right]\,,\nonumber\\
&g(\tau)=\left\{\begin{array}{ll}
\sqrt{\tau^{-1}-1}\arcsin\sqrt{\tau}
& \tau < 1\\
\frac{\sqrt{1-\tau^{-1}}}{2}\left[\ln\frac{1+\sqrt{1-\tau^{-1}}}{1-\sqrt{1-\tau^{-1}}}
-i\pi\right] & \tau \geq 1
\end{array}
\right. \, .
\end{align}

Then the ratio of decay rates is
\begin{eqnarray}
\alpha_{\gamma Z}&=&\frac{|\mathcal{M}( h\to\gamma Z)|^2}{|\mathcal{M}_{\mathrm{SM}}(h\to\gamma Z)|^2}\nonr\\
&=&\left|1+ \frac{8\sqrt{2}\pi^2 a_2}{G_F\Lambda^2(A_F+A_W)}\right|^2+
\left|\frac{8\sqrt{2}\pi^2 \tilde{a}_2}{G_F\Lambda^2(A_F+A_W)}\right|^2\, .
\end{eqnarray}

\subsection{$h\to WW^*, ZZ^*$}

Unlike the cases of $h \rightarrow gg, \gamma\gamma, \gamma Z$ which are 1-loop processes in the SM, the leading contributions
to $h\to WW^*, ZZ^*$ start at tree-level in the SM. Moreover, the SM tree-level $h V^\mu V^\nu$  vertex is $\sim g^{\mu\nu}$ which
mixes nontrivially with the form factor $S^{\mu\nu}$ when squaring the amplitude.
So $\alpha_{WW,ZZ}$ can not be expressed as $|1+\epsilon c_{even}|^2+|\epsilon' c_{odd}|^2$, where $\epsilon,\epsilon'$ are some small numbers, as in the previous cases.
To obtain the $h\to V V^*$(V=W,Z) decay, one needs to perform a straightforward tree-level calculation
and take care of the phase space integration of the 3-body final state.  When the decay final states of gauge boson $V^*$
are massless, the expression can be largely simplified but still not very illustrating. The details can be found in the appendix.
With the presence of the gauge-Higgs operators, the ratio of $h\to V V^*$ decay rates are
\begin{align}
&\alpha_{WW}=\frac{R_1(m_W/m_h, a_W)}{R_1(m_W/m_h, 0)}+\tilde{c}^2_2
\frac{32 m_h^2m_W^2}{\Lambda^4}\frac{R_2(m_W/m_h)}{R_1(m_W/m_h, 0)}\,,
\nonumber\\
&\alpha_{ZZ}=\frac{R_1(m_Z/m_h, a_Z)}{R_1(m_Z/m_h, 0)}+\tilde{a}^2_4
\frac{32 m_h^2m_Z^2c_W^4}{\Lambda^4}\frac{R_2(m_Z/m_h)}{R_1(m_Z/m_h, 0)}\,,
\end{align}
where the relevant CP even NP is encoded by two parameters:
\begin{equation}
a_W =   \frac{8m_W^2}{\Lambda^2}c_2\nonumber\,,\;
a_Z =   \frac{8m_W^2}{\Lambda^2}a_4\,.
\end{equation}

Our results for $\alpha_{gg,\gamma\gamma, \gamma Z}$ agree with \cite{MWPLB} and  the explicit expression for $\alpha_{WW,ZZ}$ are however new.
Now we are fully equipped for the later numerical analysis on the LHC Higgs decay data.

\subsection{Numerical expressions for $\alpha$'s}
Taking $\Lambda=1$ TeV as a reference point, numerically, we have
\beqa
\label{eq:alpha_Num}
\alpha_{gg} &=& \gamma_{gg}= (1+13.90 c_3)^2\,,\nonr\\
\alpha_{\gamma\gamma} &=& [1-1.48 (c_1+c_2)]^2 +2.18(\tilde{c}_1 +\tilde{c}_2-\tilde{c}_{12})^2\,,\nonr\\
\alpha_{\gamma Z} &=& \left[1+0.46 c_1-1.54 c_2\right]^2+0.71 \left[0.55\tilde{c}_1-1.82\tilde{c}_2+0.64\tilde{c}_{12}\right]^2\,,\nonr\\
\alpha_{WW} &=&1-0.019c_2+1.3\times10^{-4}c_2^2+5.2\times 10^{-5} \tilde{c}_2^2\,,\\
\alpha_{ZZ} &=& 1-0.86\times 10^{-3}c_1-0.0096c_2+2.6\times 10^{-7}c_1^2+5.8\times10^{-6} c_1c_2+3.0\times 10^{-5}c_2^2\nonumber\\
&&+ 1.3\times 10^{-5}(\tilde{c}_2 + 0.09\tilde{c}_1+ 0.28\tilde{c}_{12})^2\,.\nonr
\eeqa
By re-scaling   $c_i\rightarrow c_i\times (\Lambda/1 \mbox{TeV})^2 $,
one can easily obtain the numerical expression for the cutoff scale which differs from $1$TeV.
Since the $h\rightarrow \gamma\gamma, Z\gamma, gg$ are the loop precesses in the SM, these modes are more sensitive to the NP.
One can see that the prefactors associated with $c$'s and $\tilde{c}$'s in $\alpha_{gg,\gamma\gamma,\gamma Z }$ are all around a fews.
On the other hand, the leading contributions for $h\rightarrow WW, ZZ$ begin at tree-level in the SM. The NP modification to these modes
are relatively two orders weaker,  by a loop factor $\sim 1/16\pi^2$, compared to $h\rightarrow gg,\gamma\gamma,\gamma Z $.

\section{EDM and g-2 }\label{EDM-calculation}

\begin{figure*}[h]
$$
\includegraphics[width=12cm]{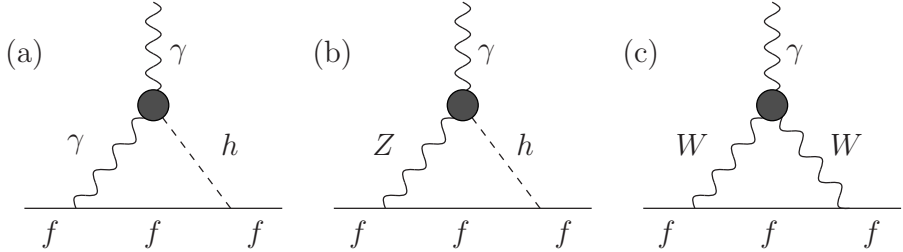}
$$
\caption{The 1-loop  contributions to electron and quark EDMs and charged lepton g-2.
The gray bulbs represent the effective gauge-Higgs operators. Note that the mirror image diagrams are not displayed. }
\label{fig:g2EDM}
\end{figure*}
First, we set our convention.
The electric dipole moment (EDM) of a fermion, $d_f$,  is defined by the low energy effective Lagrangian
\begin{equation}
\mathscr{L}_{\mathrm{EDM}}=-\frac12 d_f \bar{u}(p_2)i\gamma_5\sigma_{\mu\nu} u(p_1) F^{\mu\nu}\,,
\end{equation}
where $u$ and $\bar{u}$ are the spinor wavefunction.
It is well known that the fermion EDM starts at least at 3-loop level for quarks and 4-loop level for leptons in the SM.
However, with the CP-odd operators in the effective Lagrangian (\ref{Lan}), the fermion EDM  can be generated at one-loop level,
and there are three possible  contributions from the CP-odd $h\gamma\gamma$, $h\gamma Z$, and $\gamma WW$ interactions, see Fig.\ref{fig:g2EDM}.
We note in passing that  quarks can acquire nonvanishing  chromoEDM if $\tilde{c}_3\neq 0$.
Similar consideration of using the EDM to constrain the CP-odd Wilson coefficients  can be found in \cite{Pospelov,ZY}, where only the
contributions from Fig.\ref{fig:g2EDM}(a) has been taken into account.
As discussed in the introduction, we perform the 1-loop calculation by dimensional regularization
 and extract the finite part by $\overline{MS}$ scheme. Our result for the complete 1-loop contributions is summarized as
\beq
d_f= -e \frac{\alpha}{\pi} \frac{m_f}{v^2}\left[
Q_f  \tilde{a}_1 K_1(\Lambda,m_h) + \tilde{a}_2 \frac{\frac12I_f-Q_f s_W^2}{s_W^2 s_{2W}}K_2(\Lambda,m_Z,m_h) +
\frac{\tilde{c}_{12}}{4s_W^2} K_1(\Lambda,m_W)
\right]\,,
\eeq
where the definition of $\tilde{a}_1, \tilde{a}_2$ have been given in Eq.(\ref{eq:def_coef_a}), $I_f$ is fermion's isospin,
and in our convention, $e=|e|$ and $Q_e=-1$.
In the bracket,
the first, second, and third term
represent the contributions from Fig.\ref{fig:g2EDM}(a),(b), and (c) respectively.
In the limit that $m_f\ll m_h$, the functions $K_{1,2}$ take the form
\beqa
\label{eq:def_K_12}
 K_1(\mu, x) &\equiv& \frac{v^2}{\Lambda^2}\left[\frac34 + \frac12
\ln\frac{\mu^2}{x^2}\right]\,, \nonr\\
 K_2(\mu, x, y)& \equiv &
\frac{v^2}{\Lambda^2}\left[
\frac34+\frac12 \frac{(x^2\ln\frac{\mu^2}{x^2}-y^2\ln\frac{\mu^2}{y^2})}{(x^2-y^2)}
\right]\,,
\eeqa
where $\mu$ is the dimensional regularization scale.
Similar result for the $h\gamma\gamma$ contribution can be found
in \cite{Pospelov}, our result is different from \cite{ZY} up to a finite term.
For neutron,  we adopt the  QCD sum rule estimation
 \begin{equation}
 d_n=(1\pm 0.5 )\left[1.1 e (\tilde{d}_d+0.5 \tilde{d}_u)+1.4(d_d-0.25d_u)
 \right]\,,
 \end{equation}
 given in \cite{Pospelov-nEDM}
to relate the quark EDM and neutron EDM\footnote{ In principle, when RG running is taken into account one gets a better estimation of neutron EDM. See \cite{Xu_nEDM} for the discussion of RG running of dim-5, dim-6 CP odd operators
and the bound on the left-right model scale from the neutron EDM.
For a recent study on the RG running of dim-6 CP odd operators and the neutron EDM bound, see \cite{Dekens:2013zca}.}.
We take $m_u=2.3$MeV,  $m_d=4.8$MeV, and $\alpha=1/128.0$ for our numerical study.
 The quark chromoEDM's are ignored due to the $\theta_{QCD}$ constraint.
In terms of the Wilson coefficients, we obatin
 \begin{align}
&d_e=(7.00\tilde{c}_1+7.39\tilde{c}_2-16.07\tilde{c}_{12})\times 10^{-26}~ e\,\mathrm{ cm}\,,\nonumber\\
&d_n=(1.91\tilde{c}_1+10.04\tilde{c}_2-16.25\tilde{c}_{12})\times 10^{-25}e\, \mathrm{cm}\,.
 \end{align}
From the latest bounds: $|d_e|<1.05\times 10^{-27}$ e cm(90\% C.L.)\cite{eEDM-exp},
 and $|d_n|< 2.9\times 10^{-26}$ e cm (90\% C.L.)\cite{nEDM-exp},
we obtain two inequalities
 \begin{align}
&|66.69\tilde{c}_1+70.37\tilde{c}_2-153.01\tilde{c}_{12}|<1 \qquad (1.6 \sigma)\,,\nonumber\\
&(1\pm 0.5)\times |6.60 \tilde{c}_1+34.64\tilde{c}_2-56.02 \tilde{c}_{12}|<1  \qquad (1.6 \sigma)\,,
 \end{align}
When combining the above two conditions together, the allowed region is  a solid tube with tiny parallelogram cross section passing through the origin  along  the  $\{ 0.735, 1.478, 1.0 \}$ direction in the  $\{ \tilde{c}_1,\tilde{c}_2,\tilde{c}_{12} \}$ space.

The same 1-loop diagrams in Fig.\ref{fig:g2EDM} with CP-even operators give rise to $a_f$, the   anomalous magnetic dipole moment (AMDM) of charged leptons. The AMDM can be extracted from the charged lepton's photon form factor,
\begin{equation}
i\mathcal{M}=i e\ubar(p_2)\left[\gamma^\mu F_1(q^2)+\frac{i\sigma^{\mu\nu}q_\nu}{2m}F_2(q^2)\right]u(p_1),
\qquad
a_f\equiv\frac{g-2}{2}=F_2(0)\,,
\end{equation}
where the momentum transfer $q=p_2-p_1$.
Again, we perform the 1-loop calculation, in the unitary gauge,  with dimensional regularization and  in the $\overline{MS}$ scheme.
  We obtain the following result
\beq
a_f= -\frac{\alpha}{\pi}\frac{ m_f^2}{v^2} \left[  2 a_1 Q_f  K_1(\Lambda,m_h) +2 a_2 \frac{ \frac12 I_f-Q_f s_W^2}{s_W^2 s_{2W}} K_2(\Lambda,m_Z,m_h)+ \frac{5}{6}{c_2 v^2 \over s_W^2 \Lambda^2} \right]\,.
\eeq
The first and second term in the bracket are the finite part of  Fig.\ref{fig:g2EDM}(a) and (b) respectively.
The third term, which has no divergence,  stems from the effective dim-4 $(WW\gamma)$ interaction introduced by operator ${\cal O}_2$.
The contribution to $a_f$ from this new $(WW\gamma)$ interaction can be obtained by
multiplying the factor $-(c_2 g_2^2v^2/\Lambda^2)$ to the well known SM $WW\gamma$ contribution\cite{gm2-Wloop}.
 Numerically, we obtain $\Delta a_\mu = 1.61  ( c_1 +0.41 c_2)\times 10^{-10}$ for $\Lambda=1$ TeV.

For charged lepton anomalous magnetic diploe moment, the deviation of experimental measurement from the SM prediction are
\beqa
\Delta a_\mu &=& a_\mu^{\mathrm{exp}}-a_\mu^{\mathrm{SM}}= (2.39\pm 0.79)\times 10^{-9}\, (1\sigma)\, \mbox{ \cite{PDG2012,gm2-exp} }\,,\nonr\\
\Delta a_e &=& a_e^{\mathrm{exp}}- a_e^{\mathrm{SM}}=-10.6(8.1)\times 10^{-13}\, (1\sigma)\, \mbox{\cite{e-gm2} }\,.
\label{eq:g2bound}
\eeqa
Since the gauge-Higgs operators are flavor blind, the resulting $\Delta a_f$ scales as $m_f^2$ so $\Delta a_e = (m_e/m_\mu)^2\Delta a_\mu$.
Assuming that $\Delta a_{e,\mu}$ are solely attributed to the gauge-Higgs operators and using the latest data, Eq.(\ref{eq:g2bound}),
 we obtain a best fit at
 $\Delta a_\mu=2.37\times 10^{-9}$ from the least square fit, and
\begin{equation}
c_1+ 0.41 c_2=15.25 \pm 5.09\,.
\end{equation}
The allowed region is basically a wide infinite strip  away from the origin on the $c_1$-$c_2$ plane.

\section{Numerical analysis}

 To quantify how  each Higgs decay channel differs from its SM prediction at the LHC,
we follow \cite{width}  and use the  signal strength parameter $\hat{\mu}$ which is  defined as
\begin{equation}
\hat{\mu}_{ij}=\frac{\sigma(X\to h)\mathcal{B}(h\to ij)}{\sigma(X\to h)_{\mathrm{SM}}\mathcal{B}(h\to ij)_{\mathrm{SM}}}\,,
\end{equation}
where $X$ stands for the initial partons in proton, and $i,j$ represent the decay products.
The branching fractions are given by $ \mathcal{B}_{ij}\equiv
\mathcal{B}(h\to i j )=\Gamma_{ij}/ \Gamma_{\mathrm{tot}}$,
and $\Gamma_{\mathrm{tot}}$ is the actual total Higgs decay width. The $\Gamma_{\mathrm{tot}}$ is related to the
SM prediction by $\Gamma_{\mathrm{tot}}=C_{\mathrm{tot}}\Gamma_{\mathrm{tot}}^{\mathrm{SM}}$.
Then, in terms of $\alpha$'s  and $\mathcal{B}$'s, $C_{\mathrm{tot}}$ can be expressed as the sum of all contributions from the major decay channels
\begin{eqnarray}
C_{\mathrm{tot}}&\sim &\mathcal{B}_{bb}^{\mathrm{SM}}+\mathcal{B}_{\tau\tau}^{\mathrm{SM}}
+\mathcal{B}_{cc}^{\mathrm{SM}}+\alpha_{\gamma\gamma}\mathcal{B}_{\gamma\gamma}^{\mathrm{SM}}
+\alpha_{gg}\mathcal{B}_{gg}^{\mathrm{SM}}+\alpha_{WW}\mathcal{B}_{WW}^{\mathrm{SM}}+
\alpha_{ZZ}\mathcal{B}_{ZZ}^{\mathrm{SM}}+
\alpha_{Z\gamma}\mathcal{B}_{Z\gamma}^{\mathrm{SM}}\nonumber\\
&\sim & 0.67+0.0023\alpha_{\gamma\gamma} +0.086\gamma_{gg}+0.026\alpha_{ZZ} +0.215 \alpha_{WW}
+0.0015\alpha_{Z\gamma}\,,
\label{eq:C_tot}
\end{eqnarray}
where the SM branching ratios for a 125 GeV Higgs are adopted from \cite{Higgs-Xsec} and note that  $\alpha_{ff}=1$
at tree-level since the gauge-Higgs operators do not modify the Yukawa interactions.
Due to the parton distribution function, the loop induced gluon fusion is the dominate production mechanism ($\sim 87-88\%$ )
at the LHC for a Higgs with mass around $125$ GeV.
And among all production channels, it is most sensitive to new physics. Therefore, it is a fairly good approximation
to take $\sigma(gg\to h)/ \sigma(gg\to h)_{\mathrm{SM}} \sim \gamma_{gg}$.
Therefore we have
\begin{equation}
\hat{\mu}_{ij} \sim \frac{\gamma_{gg}\alpha_{ij}}{C_{\mathrm{tot}}}\,.
\end{equation}
As the common practice,  a function $\chi^2$ is defined for multi-parameters fitting
\begin{equation}
\chi^2= \sum_i \frac{(\hat{\mu}_i -\bar{\mu}_i)^2}{\delta_i^2}\,,
\end{equation}
where $\bar{\mu}_i$ is the mean experimental value, $\delta_i$ is the combined uncertainty from both the experimental and theoretical sides, and $\hat{\mu}_i$ is the theoretic prediction either from a specific model or determined by the effective gauge-Higgs operators. We use the most updated Higgs data collected and analyzed
by ATLAS and CMS groups, see Tab.\ref{tab:Higgs-Moriond13}.
\begin{table}[h]
\renewcommand{\arraystretch}{1.2}\addtolength{\arraycolsep}{3pt}
\caption[]{The signal strength $\bar{\mu}$ from Moriond 2013. }
$$
\begin{array}{c|c|c}
\hline
\bar{\mu}_{ij} & \mathrm{CMS} &  \mathrm{ATLAS} \\
\hline\hline
 ZZ^*       &0.91^{+0.30}_{-0.24} ~(\mathrm{inclusive})     &  1.7^{+0.5}_{-0.4}  ~(\mathrm{inclusive})
 ~\mbox{ \cite{ATLAS-ZZ,ATLAS-bosonic}   }   \\
            &1.0^{+2.4}_{-2.3}~ (\mathrm{qqH,VH}) &   \\
            &0.9^{+0.5}_{-0.4} ~(\mathrm{GF}) ~~\mbox{\cite{CMS-bosonic, CMS-ZZ} }  &         \\
 WW^*   & 0.76^{+0.21}_{-0.21}  ~(\mathrm{inclusive }) ~~\mbox{\cite{CMS-bosonic, CMS-WW} }
 & 1.01\pm0.31 ~(\mathrm{inclusive}) ~ \mbox{\cite{ATLAS-bosonic-QCD, ATLAS-WW}} \\
 \gamma\gamma &0.78^{+0.28}_{-0.26}~(\mathrm{MVA})    & 1.65^{+0.34}_{-0.30}   ~(\mathrm{inclusive})
 ~\mbox{\cite{ATLAS-gamma, ATLAS-bosonic} }     \\
  & 1.11^{+0.32}_{-0.30}  ~(\mathrm{cut\;based})~\mbox{\cite{CMS-bosonic-QCD, CMS-gamma} } &     \\
  \gamma Z & <9  ~\mbox{\cite{CMS-gammaZ}}  &  <18.2 ~\mbox{\cite{ATLAS-gammaZ, ATLAS-bosonic}} \\
 \hline
 bb & \mathrm{HCP12}:~~ 1.3^{+0.7}_{-0.6} ~(\mathrm{VH})~ \mbox{\cite{LHC-fermionic} } & 1.09\pm0.20\pm0.22
 ~(\mathrm{VH})~\mbox{\cite{ATLAS-fermionic, ATLAS-bb} }      \\
  \tau\tau &1.1^{+0.4}_{-0.4} ~(\mathrm{inclusive})~ \mbox{\cite{CMS-fermionic, CMS-tautau} } &
   0.7\pm 0.7  ~(\mathrm{inclusive})  ~\mbox{\cite{LHC-fermionic, ATLAS-tautau} } \\
 \hline\hline
\end{array}
$$
\label{tab:Higgs-Moriond13}
\end{table}
Since we explicitly take $\sigma(gg\to h)/ \sigma(gg\to h)_{\mathrm{SM}} \sim \gamma_{gg}=\alpha_{gg}$,  in the CMS $h\rightarrow ZZ^*$ channels, we only use the  gluon-gluon fusion data.
 We drop $\bar{\mu}_{bb}$ from the global fitting for both groups have large uncertainties and both are via the VH production.

\subsection{Pseudo Global fit of ATLAS and CMS results }
\begin{figure*}[h]
$$
\begin{array}{cccc}
\includegraphics[width=3.5cm]{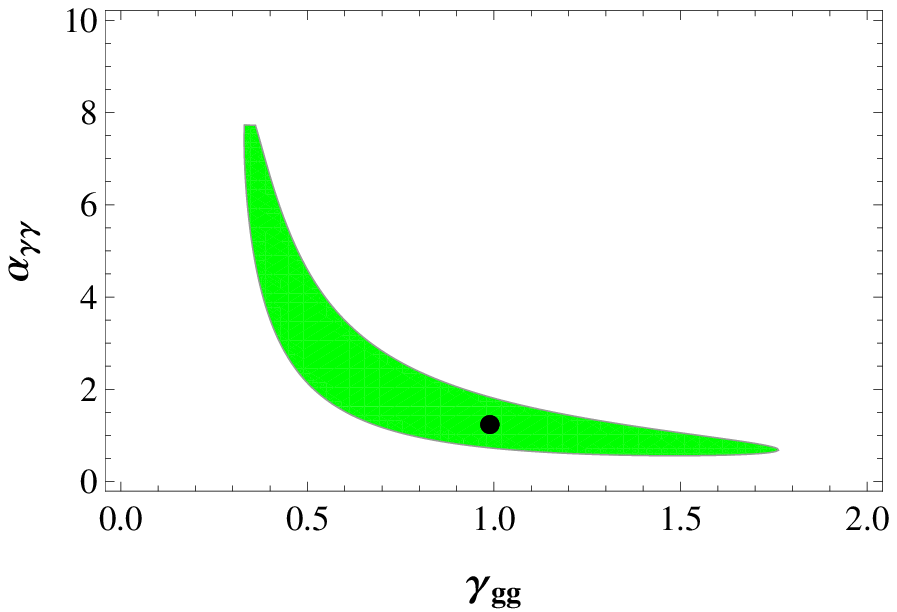}&
\includegraphics[width=3.5cm]{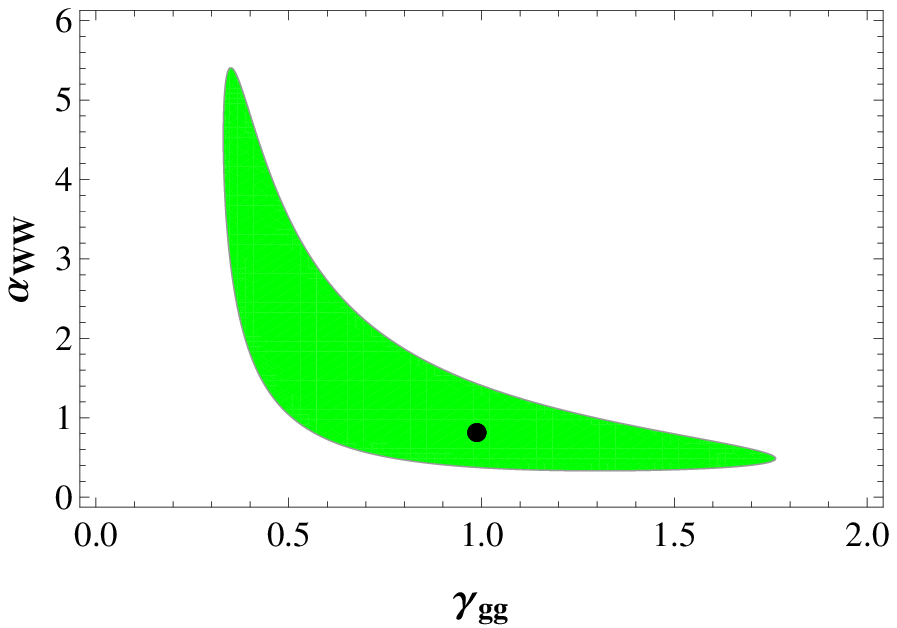}&
\includegraphics[width=3.5cm]{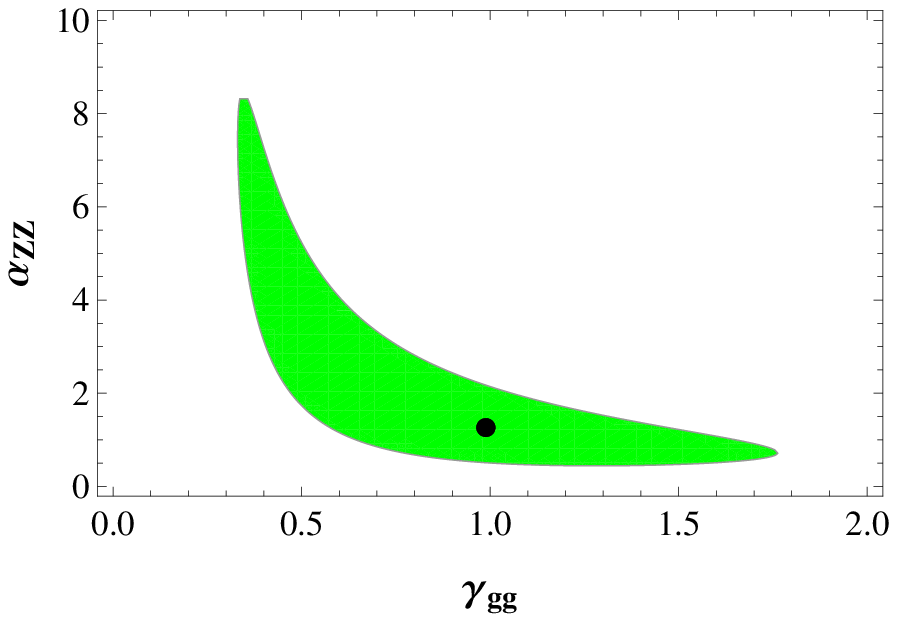}&
\includegraphics[width=3.8cm]{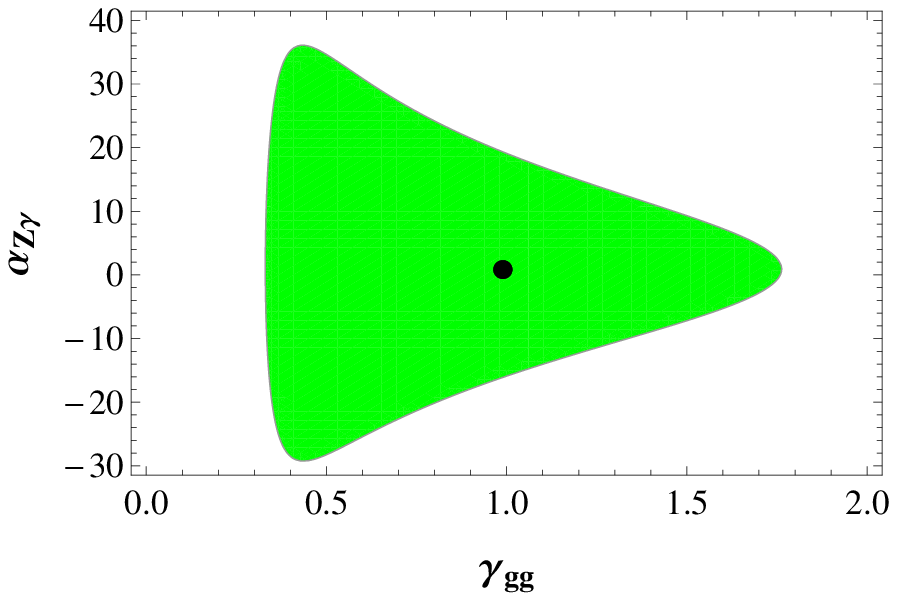}\\
(a) & (b) & (c) & (d)\\
\end{array}
$$
\caption{The $1\sigma$ correlations  between: (a) $\alpha_{\gamma\gamma}$ and $\gamma_{gg}$,
(b) $\alpha_{WW}$ and $\gamma_{gg}$,
(c)$\alpha_{ZZ}$ and $\gamma_{gg}$,
(d)$\alpha_{\gamma Z}$ and $\gamma_{gg}$. The best fit locations are marked by dots.  }
\label{fig:alpha-gamma}
\end{figure*}

To proceed, we first treat the five $\alpha$'s in Eq.(\ref{eq:C_tot}) as mutually independent free parameters and conduct a pseudo global fit by using the current LHC data listed in Tab.\ref{tab:Higgs-Moriond13}.
The best fit locates at
\begin{equation}
\gamma^{\mathrm{LHC}}_{gg}=0.99,\quad \alpha^{\mathrm{LHC}}_{\gamma\gamma}=1.26,\quad
\alpha^{\mathrm{LHC}}_{WW}=0.84,\quad
\alpha^{\mathrm{LHC}}_{ZZ}= 1.30,\quad
\alpha^{\mathrm{LHC}}_{Z\gamma}= 1.00
\end{equation}
with minimum $\chi_{0}^2=4.94$, and the $1\sigma$ boundary corresponds to the contour of $\chi_{1\sigma}^2=\chi_0^2+ 5.89 $ for a five parameters fit. And the $1\sigma$ model-independent correlations between  $\alpha_{ij}$
 and $\gamma_{gg}$ are displayed in Fig \ref{fig:alpha-gamma}.
Basically the message from doing this trivial exercise is clear and simple: (1) the current LHC data more or less agrees with the SM,
and (2) the larger the Higgs gluon-gluon fusion production  the smaller the Higgs decay widthes and vice versa. We see that the global fit prefers a SM-like  Higgs gluon fusion production. However, this still introduces a degenerate solution to $c_3$, $( 1+13.92 c_3 )= \pm \sqrt{1.0 \pm 0.7} $. And the two corresponding $1\sigma$ allowed region for $c_3$ are $\in [-0.167,-0.114]$ and $\in [-0.029,0.023]$. This degeneracy for $c_3$ also inevitably exists in the later  global fitting.  We will discuss how to lift this degeneracy phenomenologically in Sec.6.

\subsection{Limits on the gauge-Higgs Wilson coefficient by using the Higgs data alone}
\begin{figure*}[h]
$$
\begin{array}{ccc}
\includegraphics[width=5cm]{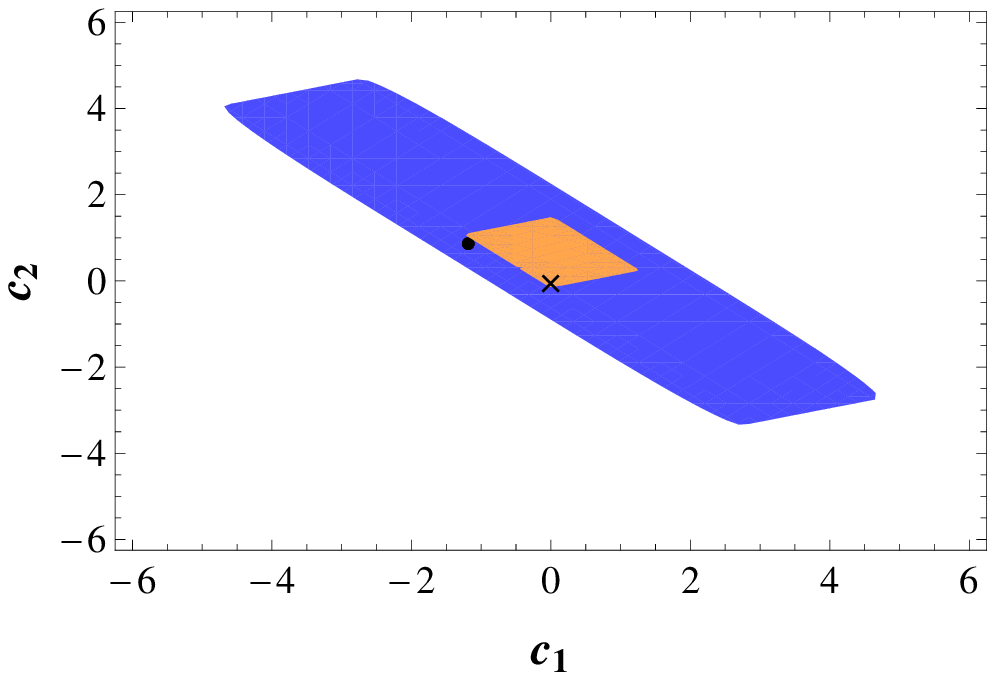}&
\includegraphics[width=5cm]{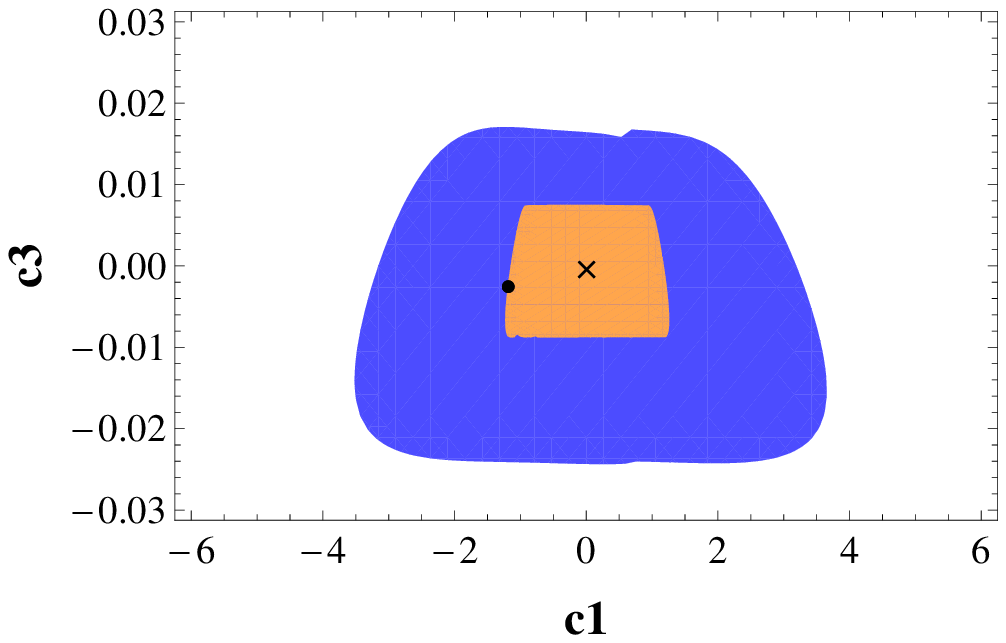}&
\includegraphics[width=5cm]{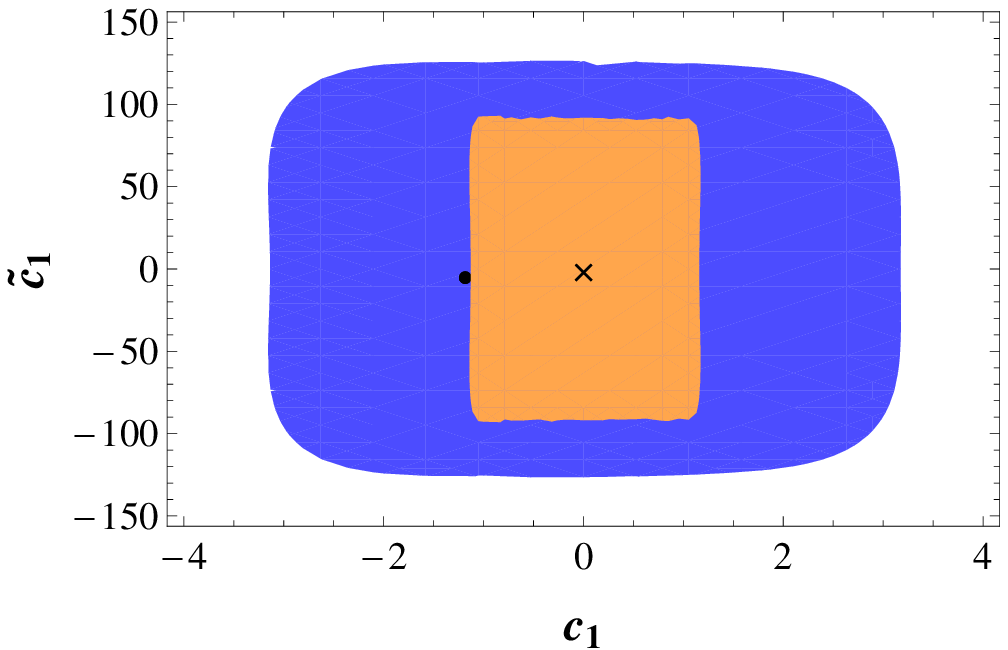}\\
(a) & (b) & (c)\\
\includegraphics[width=5cm]{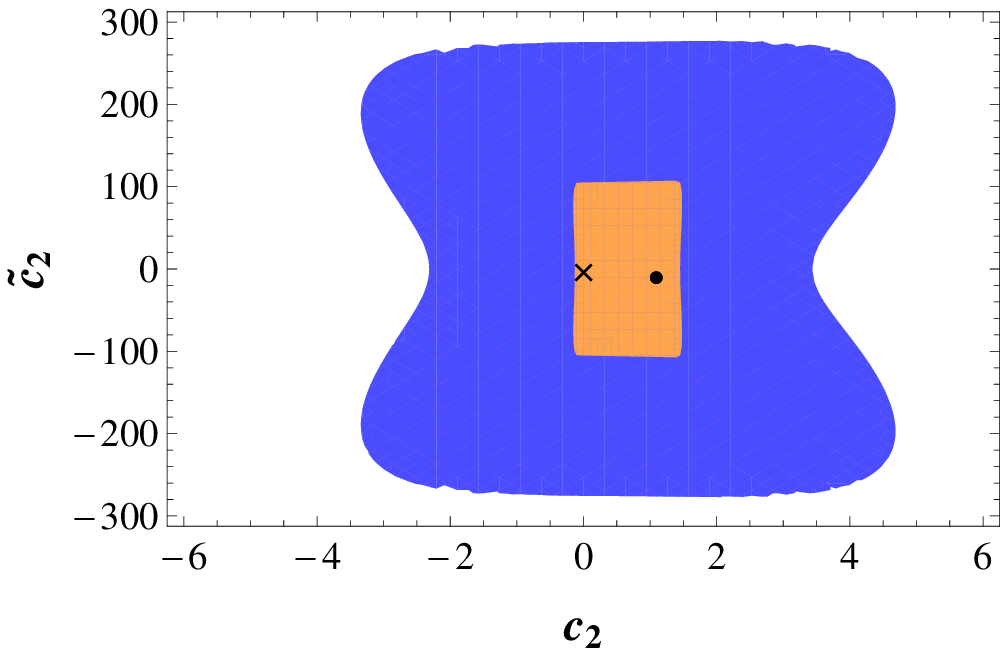}&
\includegraphics[width=5cm]{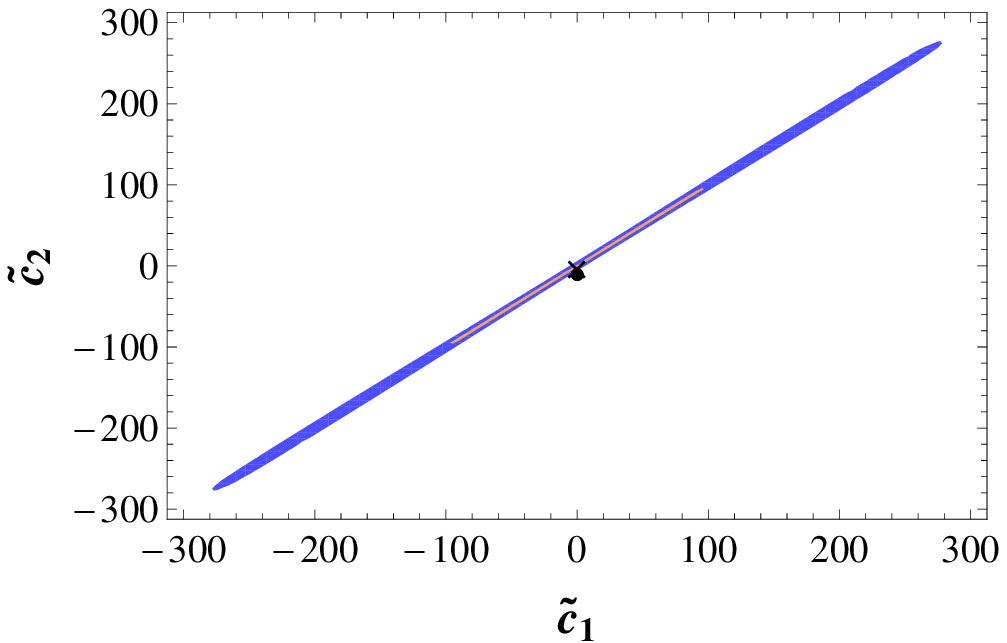}&
\includegraphics[width=5cm]{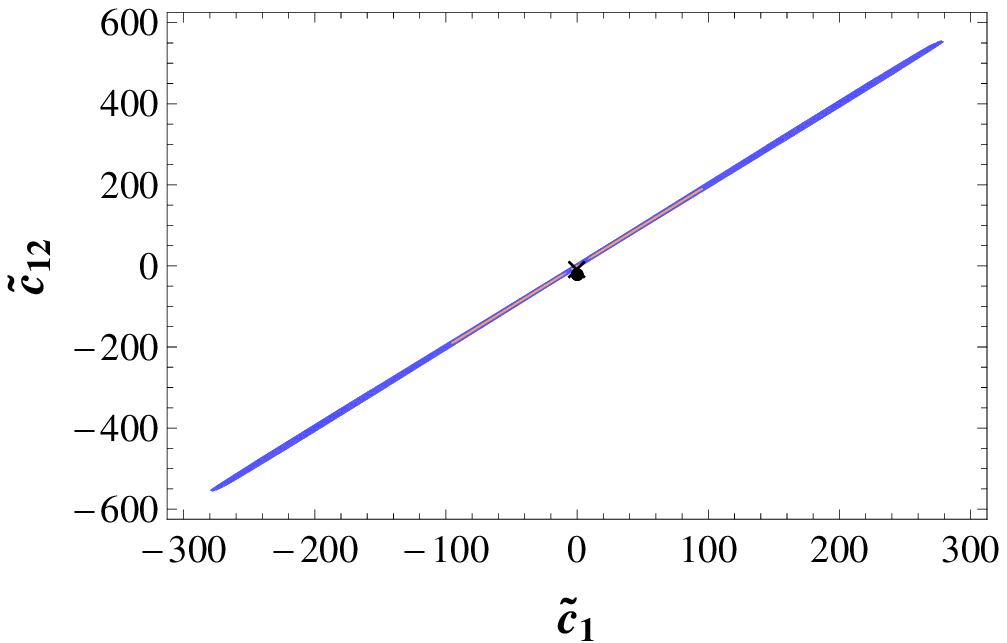}\\
(d) & (e) & (f)\\
\includegraphics[width=5cm]{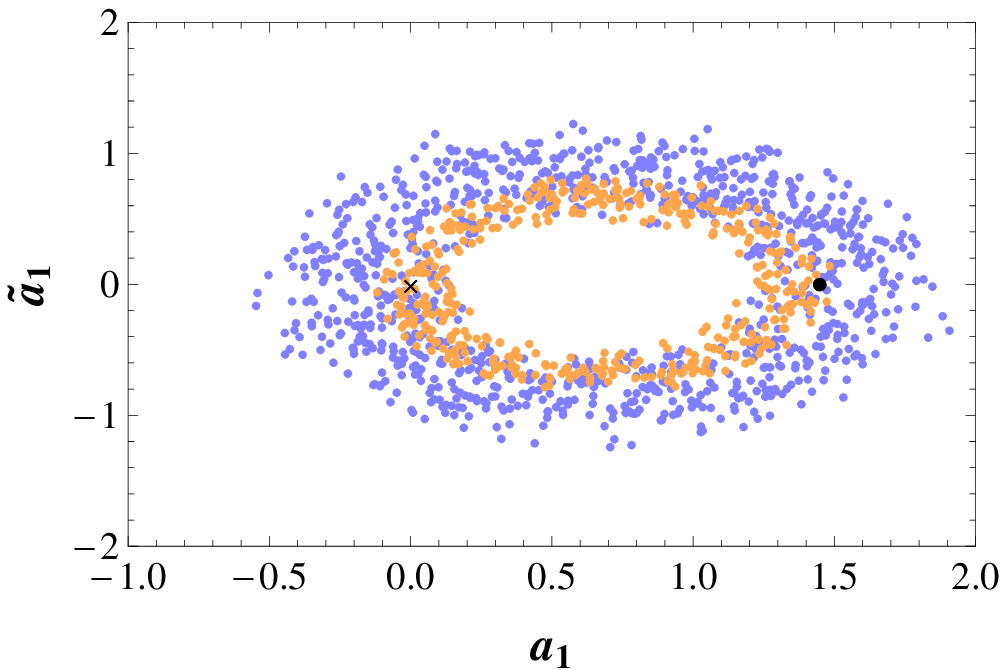}&
\includegraphics[width=5cm]{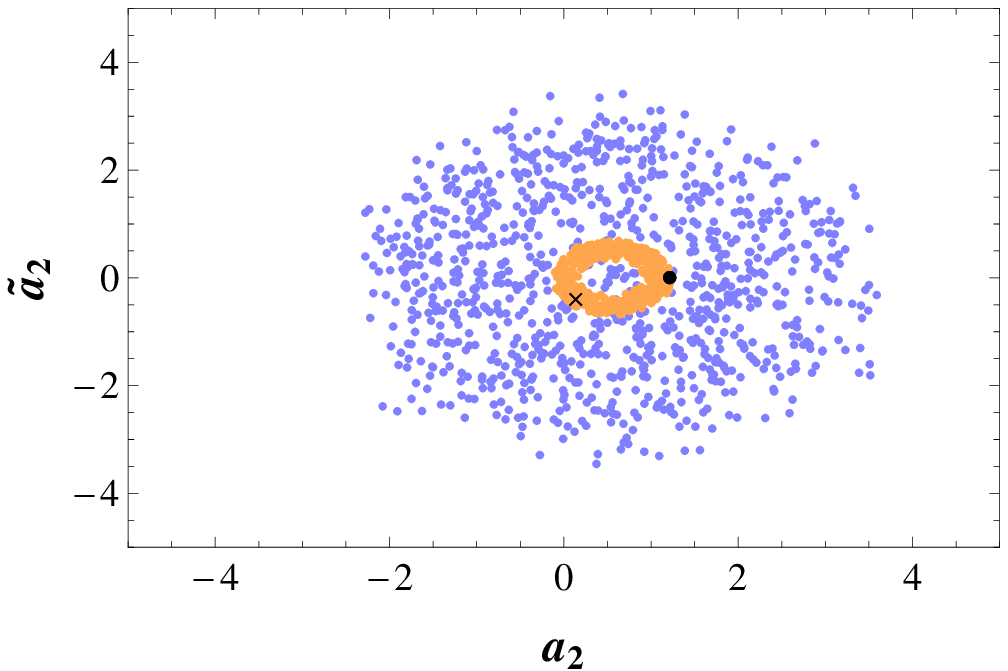}&
\includegraphics[width=5cm]{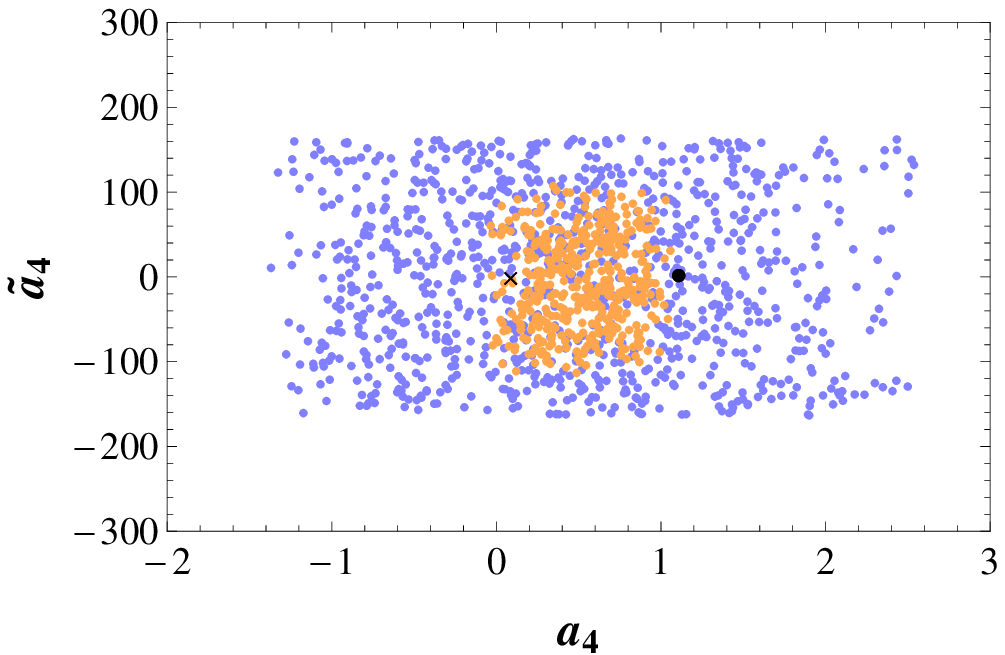}\\
(g) & (h) & (i)
\end{array}
$$
\caption{The $95\%$CL allowed region in (a-f) the Wilson coefficient space, and (g-i) the $hVV$ couplings, by using the LHC data (blue/darker)  and
the SM predictions(light brown/lighter). The best fit location is shown at the dot(cross) for LHC data (SM).
The minimum $\chi^2 = 6.413 (0.0)$ for LHC data(SM).
In subdiagram(b), we only display the SM like allowed region for $c_3$.}
\label{fig:Higgs_allowed}
\end{figure*}

Next, we do the global fit by using the current LHC data in terms of $c_{1,2,3}$ and $\tilde{c}_{1,2,12}$.
For a 6-dimensional parameter fitting, the $95\%$CL ($68\%$CL) contour corresponds to $\chi^2_{95(68)}=\chi^2_{min}+12.59(7.01)$.
 To have an idea how the Wilson coefficients will be constrained when the experimental sensitivity are improved and eventually compatible with the SM theoretical uncertainties, we also consider a `fake' data set that all signal strengthes equal one, $\bar{\mu}=1.0$. And we take the combined SM theoretical uncertainties, $\sigma_{SM} =\sqrt{\sigma_{BR}^2+\sigma_{GF}^2}$, from  Tab.\ref{tb:SM-xsec} and Tab.\ref{tb:SM-br}, as the `experimental' errors.
The 95\%CL results are shown in Fig.\ref{fig:Higgs_allowed}.
Some features of our results:
(1) The CP even Wilson coefficients are well constrained by the current LHC data.
The LHC limits on $c_1, c_2$ and $c_3$  are basically  compatible, around two times bigger, with the benchmark experimental sensitivity
by using the SM theoretical predictions and uncertainties as input.
(2) On the other hand, the constraints on the CP-odd Wilson coefficients $\tilde{c}_{1,2,12}$ are poor by using the Higgs decays data alone.
However, they seem to fall on a long line segment in the parameter space.
This linear relation can be easily understood as  roughly the solution to make the CP odd contributions vanish simultaneously
in $\alpha_{\gamma\gamma}$ and $\alpha_{\gamma Z}$, Eq.(\ref{eq:alpha_Num}), since the two are most sensitive to the presence
of CP-odd contributions.

The most important message from this drill is: there still is the allowance for NP with $c_{1,2}\sim $ a few and $\tilde{c}$'s $\sim {\cal O}(100)$  for $\Lambda=1$TeV hiding in the SM theoretical uncertainties.

\begin{figure*}[h]
$$
\begin{array}{cc}
\includegraphics[width=7cm]{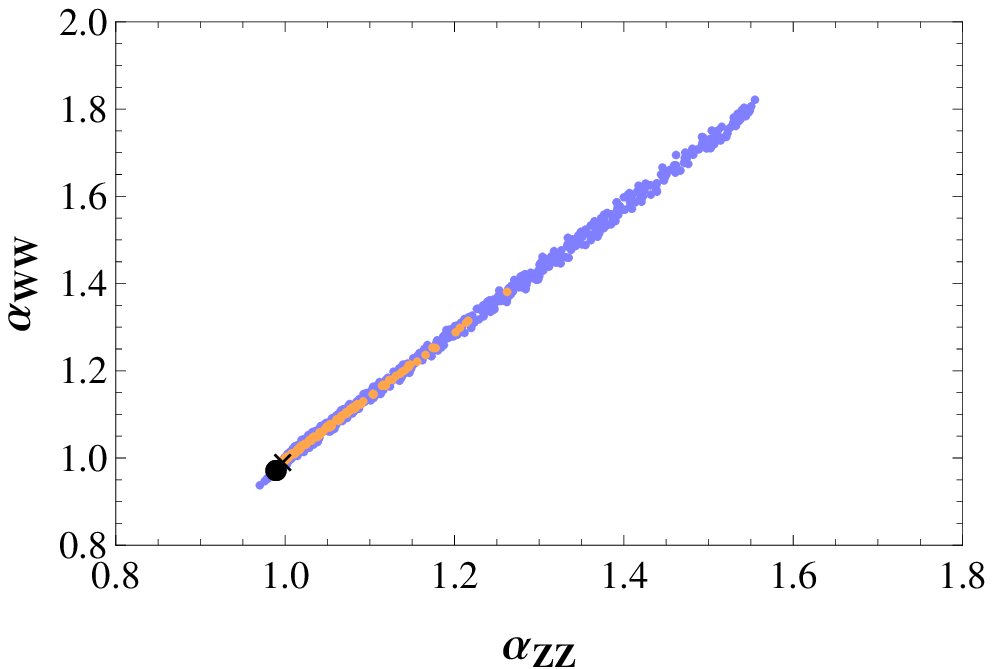}&
\includegraphics[width=7cm]{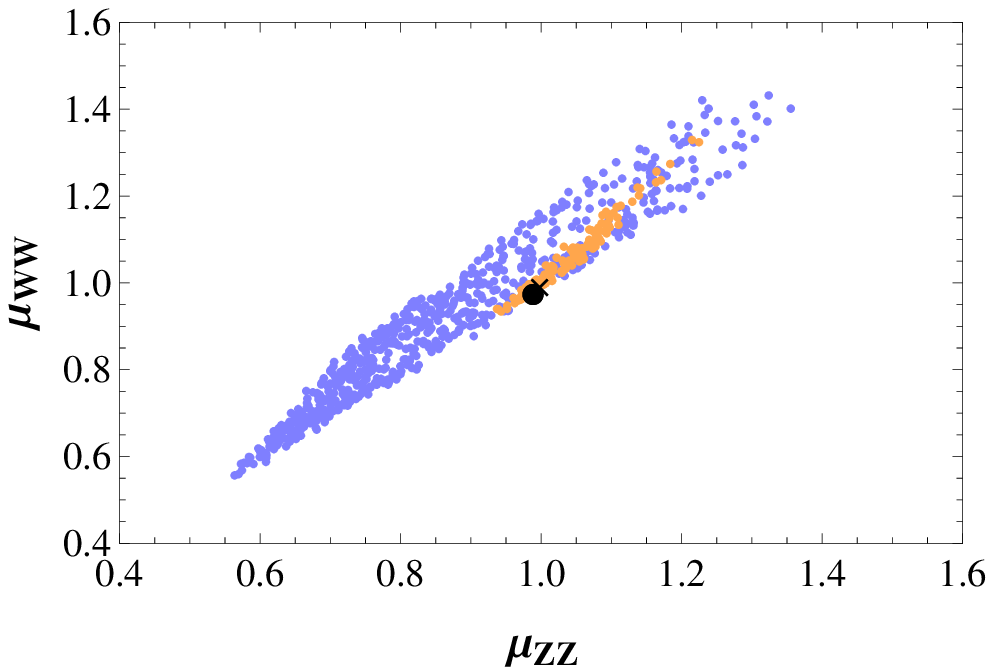}\\
(a) & (b)  \\
\end{array}
$$
\caption{The $95\%$CL correlations between the of $h\to W W^*$ and $ h\to Z Z^*$ decay rate ratios, sub-diagram (a), and
signal strength, sub-diagram (b).  Where the best fit point is marked by the dot/cross,
 and the blue (darker)/ brown(lighter) region is for using the LHC/SM data. }
\label{fig:alpha_Mu_corl_Higgs}
\end{figure*}

Interestingly, from our numerical study, we find a linear correlation among $\alpha_{WW}$ vs $\alpha_{ZZ}$ and
the signal strength $\mu_{WW}$ vs $\mu_{ZZ}$, see  Fig.\ref{fig:alpha_Mu_corl_Higgs}.
Our results show that at $95\%$ CL  $\alpha_{ZZ}\sim 1.5 \alpha_{WW}$  and $1.0 \lesssim \alpha_{ZZ}\lesssim 1.6$.
However when converted to signal strength it becomes  $\mu_{ZZ}\sim \mu_{WW}$  and $0.6 \lesssim \mu_{ZZ,WW}\lesssim 1.4$.
The best fit values for both cases are SM-like.

\subsection{Constraints on the gauge-Higgs Wilson coefficient by using the Higgs data plus EMD's and AMDMs}
\begin{figure*}[h]
$$
\begin{array}{ccc}
\includegraphics[width=5cm]{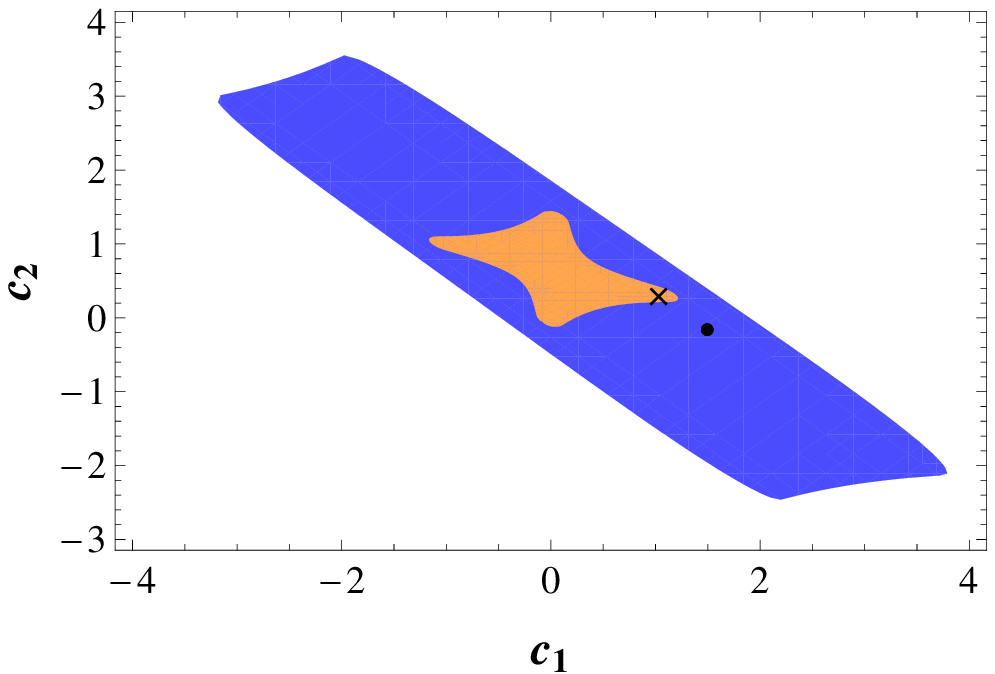}&
\includegraphics[width=5cm]{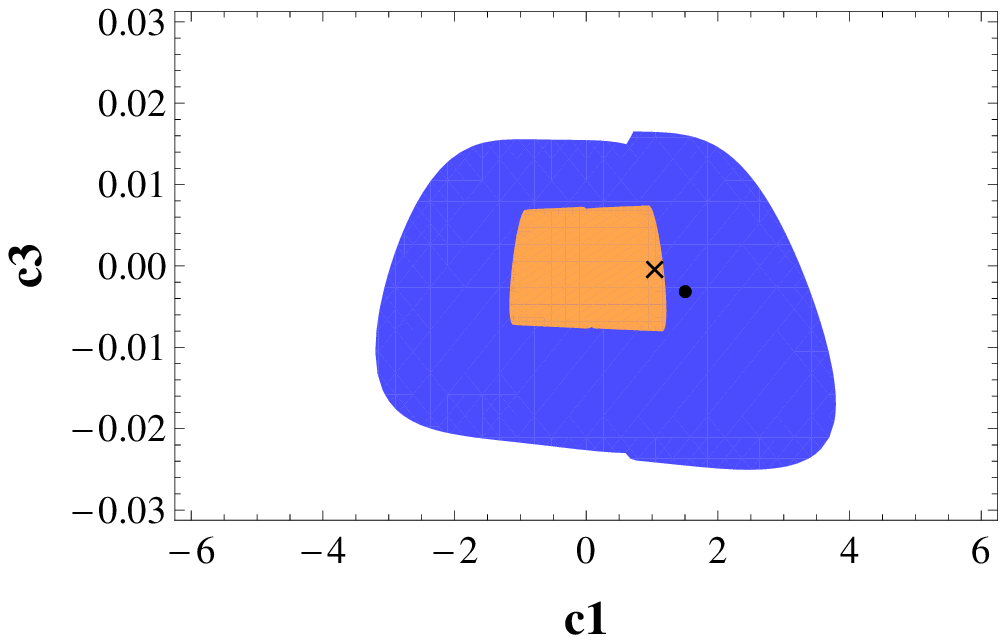}&
\includegraphics[width=5cm]{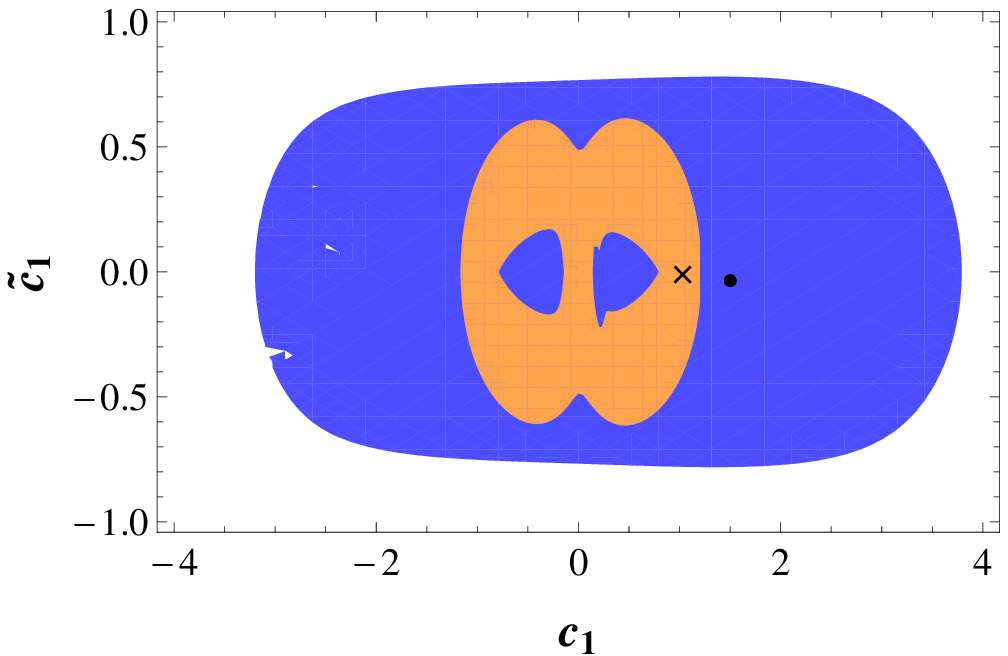}\\
(a) & (b) & (c)\\
\includegraphics[width=5cm]{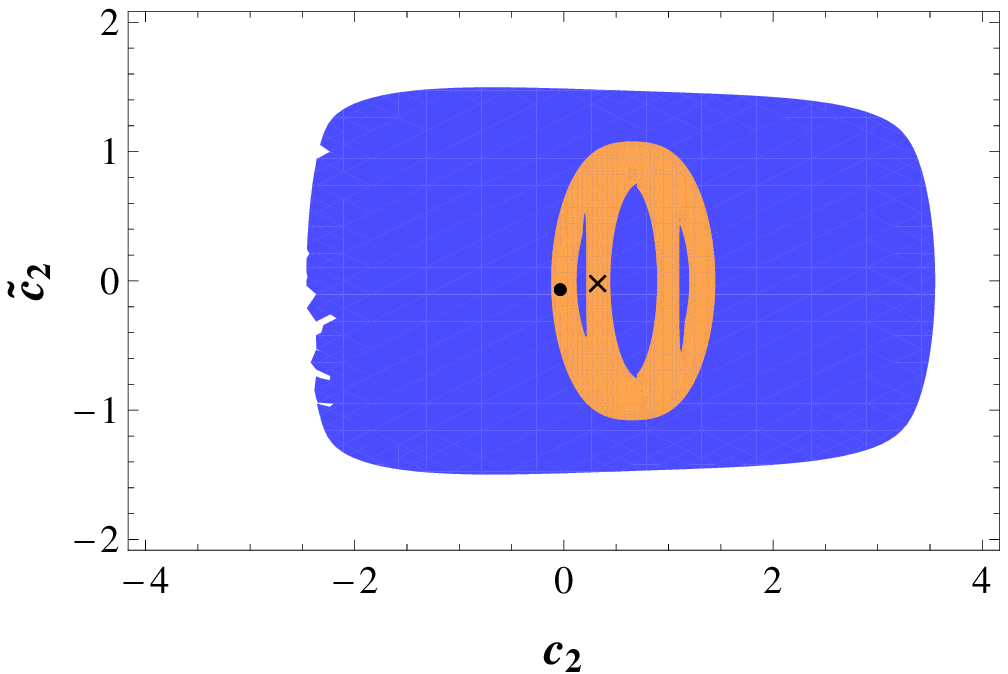}&
\includegraphics[width=5cm]{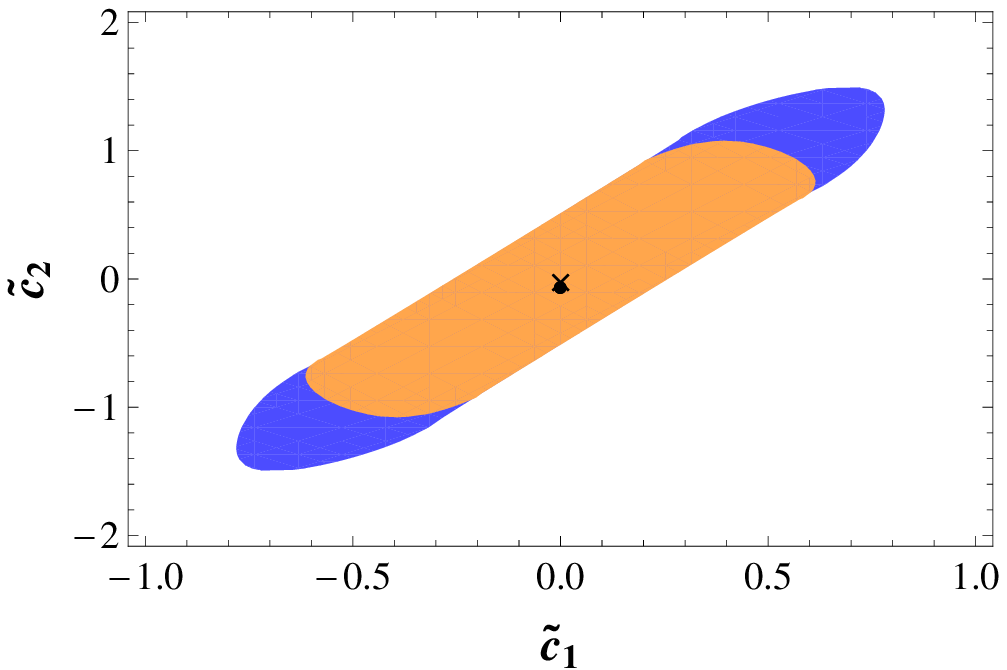}&
\includegraphics[width=5cm]{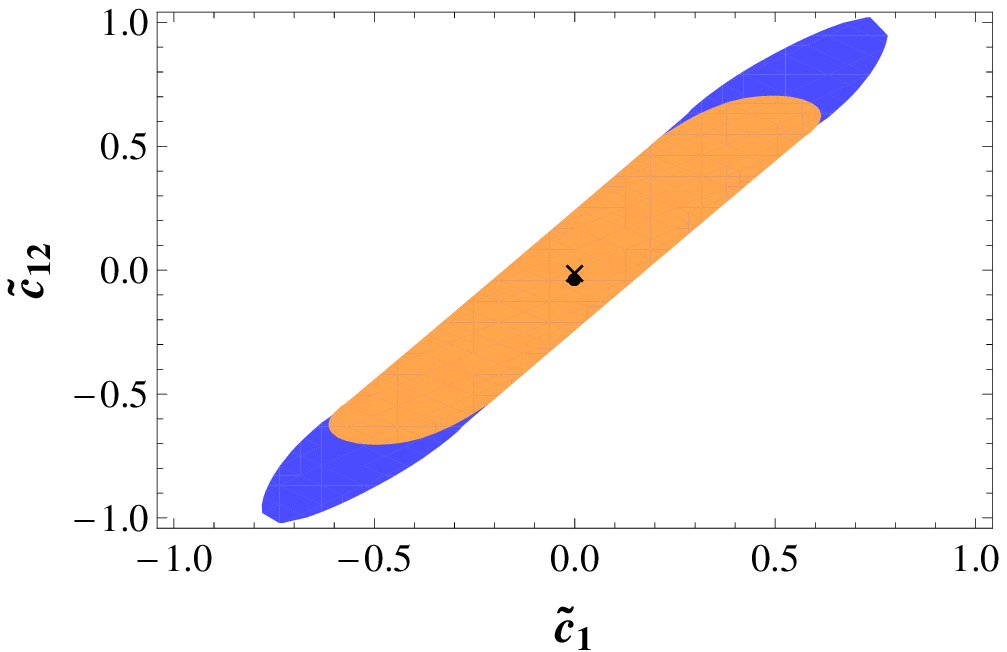}\\
(d) & (e) & (f)\\
\includegraphics[width=5cm]{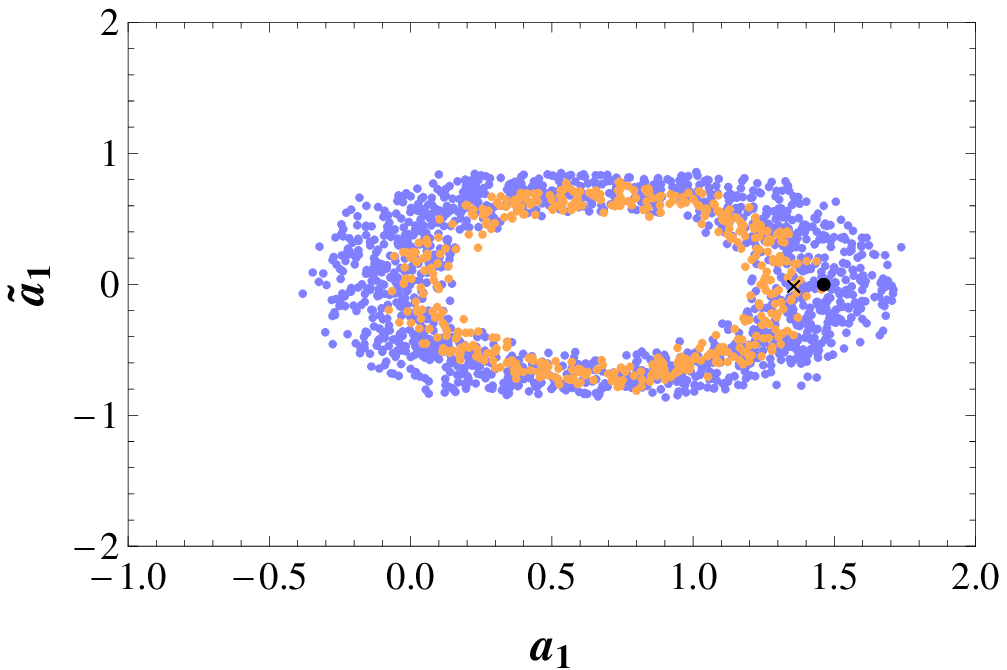}&
\includegraphics[width=5cm]{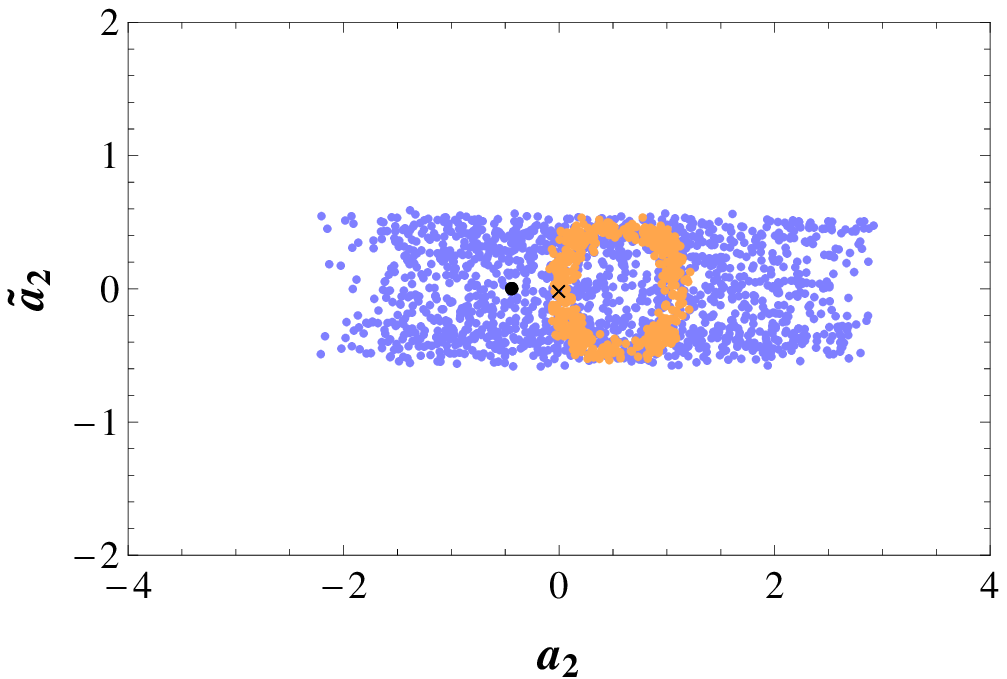}&
\includegraphics[width=5cm]{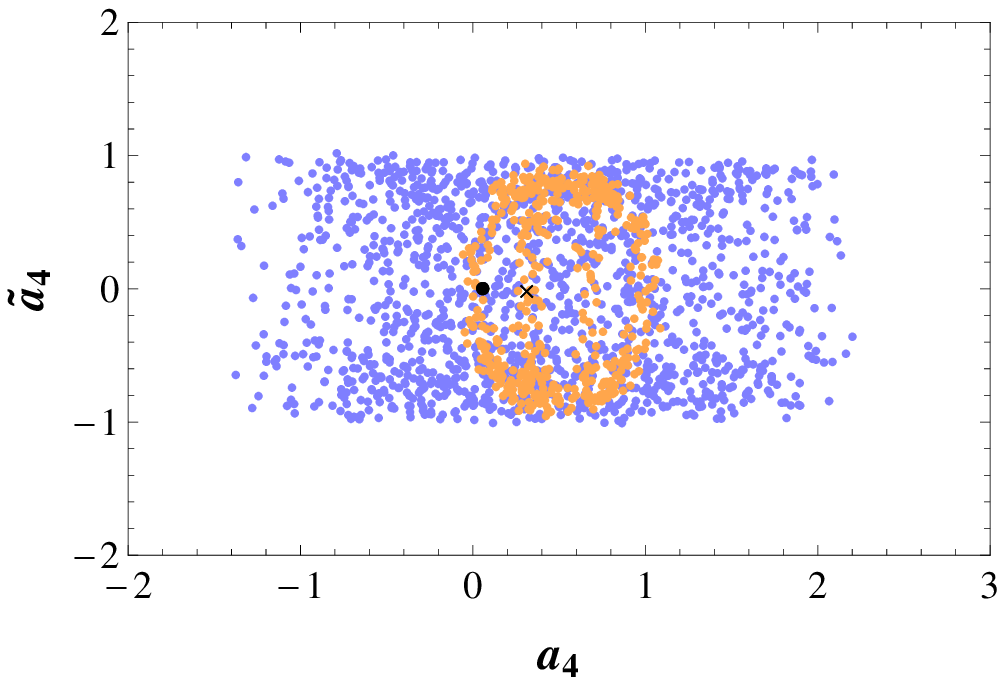}\\
(g) & (h) & (i)
\end{array}
$$
\caption{The $95\%$CL allowed region in (a-f) the Wilson coefficient space, and (g-i) in the Higgs-gauge couplings space, by using Higgs data, EDM, and AMDM.
The blue(darker) and light brown(lighter) regions are for using the LHC data  and the SM predictions respectively.
 The best fit location is shown at the dot(cross) for LHC data (SM).
The corresponding minimum $\chi^2 = 15.89 (9.63)$ for LHC data(SM).
In sub diagram (b), only the SM-like $c_3$ region is shown.}
\label{fig:Higgs_EDM_MDM_allowed}
\end{figure*}

Finally, we include 4 more data points, $d_e$, $d_n$, $\triangle a_e$, and $\triangle a_\mu$ into the global fitting.
And the $95\%$CL results are shown in Fig.\ref{fig:Higgs_EDM_MDM_allowed}.
One can clearly see that when the electron and neutron EDM are included in the global fitting, the limits on the CP-odd Wilson coefficients
are dramatically improved. Even giving extra, say $\sim {\cal O}(10)$,  uncertainties to the 1-loop EDM estimations  by using the effective operators, the inclusion of EDMs into the analysis still plays an important role to limit the CP properties of 125 GeV boson. More implications of including the EDM constraints will be discussed in Section \ref{sec:CPodd_prediction}.

\begin{figure*}[h]
$$
\begin{array}{cc}
\includegraphics[width=7cm]{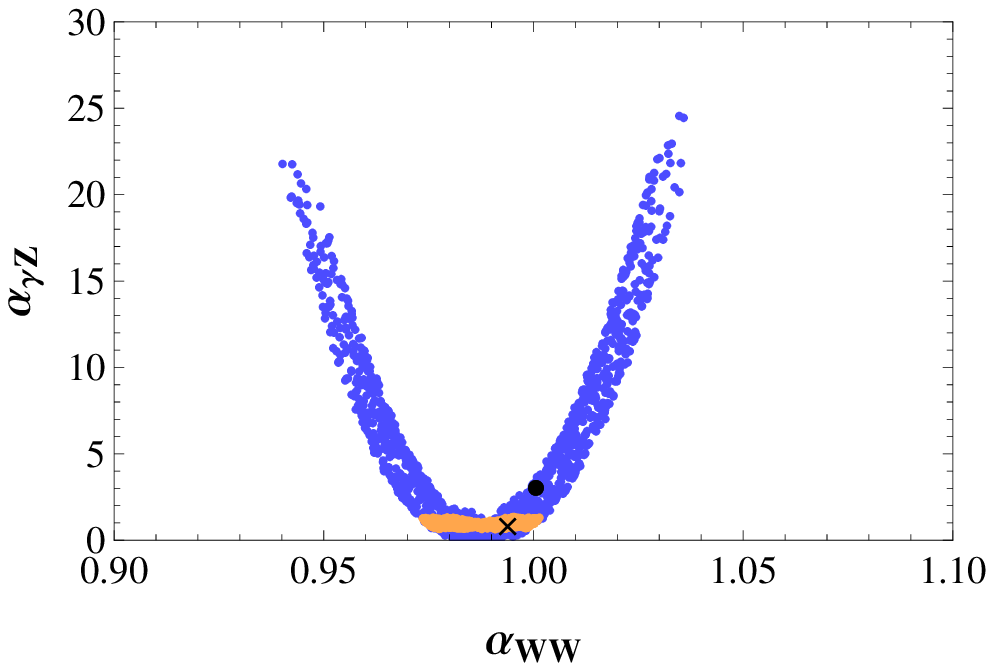}&
\includegraphics[width=7cm]{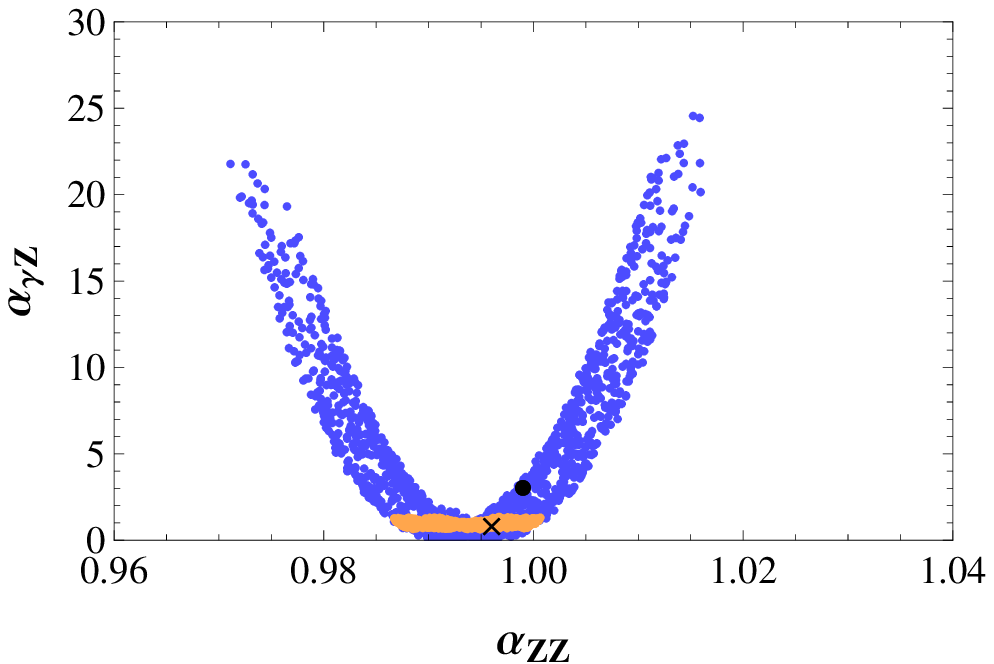}\\
(a) & (b)\\
\end{array}
$$
\caption{ The $95\%$CL decay rate ratios correlation between: (a) $h\to W W$ and $ h\to \gamma Z$ , and (b) $h\to ZZ$ and $ h\to \gamma Z$.   Where the best fit point is marked by the dot/cross,
 and the blue (darker)/ brown(lighter) region is for using the LHC/SM data. }
\label{fig:alpha_ww_zz_az}
\end{figure*}

\begin{figure*}[h]
$$
\begin{array}{cc}
\includegraphics[width=7cm]{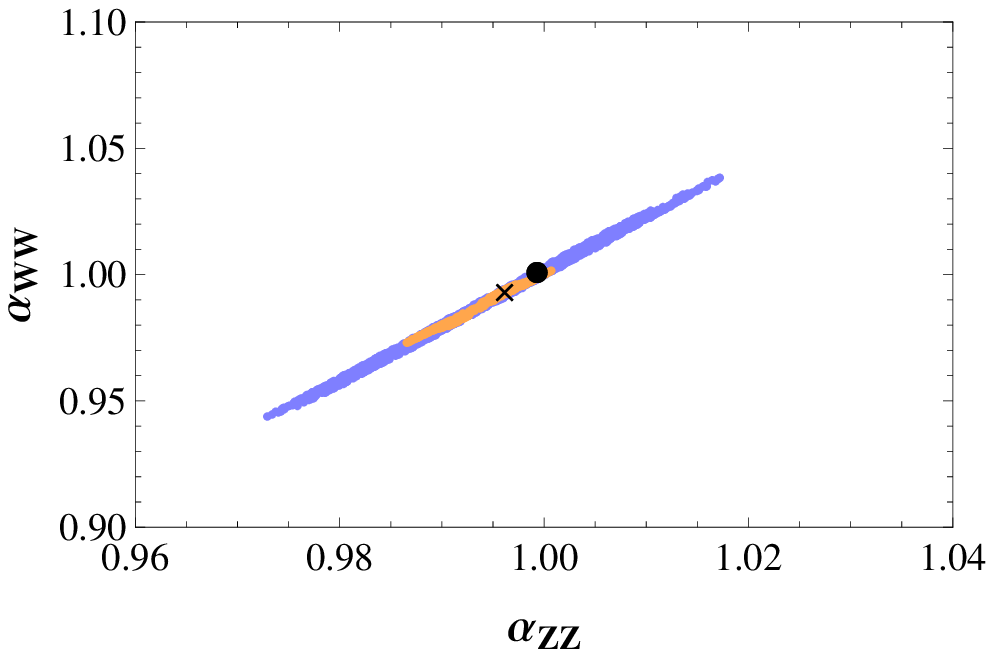}&
\includegraphics[width=7cm]{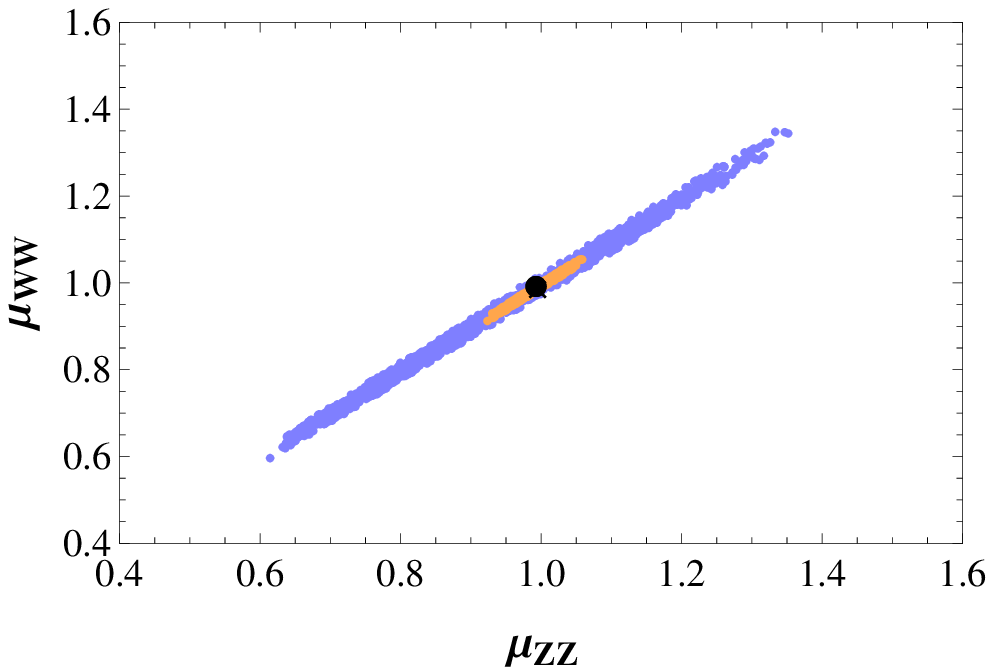}\\
(a) & (b)\\
\end{array}
$$
\caption{ This plot is similar to Fig.4 but with additional constraints from EDMs and AMDMs. }
\label{fig:alpha_correlation}
\end{figure*}

Some interesting correlations emerge among $\alpha_{\gamma Z, ZZ,WW}$, see Fig.\ref{fig:alpha_ww_zz_az} and Fig.\ref{fig:alpha_correlation}(a).
Roughly, we observe that
\beqa
&&(\alpha_{WW}-1)\sim 2 (\alpha_{ZZ}-1)\,,\nonr\\
&&(\alpha_{\gamma Z}-1)  \sim  6.5\times 10^{3} (\alpha_{WW}-1)^2 \sim  2.57\times 10^{4} (\alpha_{ZZ}-1)^2\,.
\label{eq:parabola}
\eeqa
These relations can be easily understood. Since the EDMs strongly constrain the CP-odd Wilson coefficients, the $\alpha$'s
will be dominated by the CP-even Wilson coefficients $c_1$ and $c_2$. From Eq.(\ref{eq:alpha_Num}), we see that $\alpha_{\gamma Z, WW, ZZ}$ are more sensitive to $c_2$ than $c_1$, thus the approximate relations follow.
However, when translated into the signal strength, only the relation between $\mu_{ZZ}$ and $\mu_{WW}$ is hold, see Fig.\ref{fig:alpha_correlation}(b).
Therefore, we find  that  $\mu_{ZZ}\sim \mu_{WW}$  and $0.6 \lesssim \mu_{ZZ,WW}\lesssim 1.4$ is
quite robust with or without taking EDM constraints into account.
To test the relations given in Eq.(\ref{eq:parabola}), better than $\lesssim 1\%$ accuracy of determination the individual absolute Higgs decay width is required. This will be very challenging at the LHC but could be done at the precision Higgs machines in the future.

\section{Discussion and Conclusion}

\subsection{Discriminate the degenerate solutions}
From the global fit of the latest LHC Higgs data, in the presence of effective gauge-Higgs operators, we obtain $\gamma_{gg}\simeq 1.0\pm 0.4$. This introduces a degenerate solution to $c_3$,
$ 1+13.92 c_3 = \pm \sqrt{1.0 \pm 0.4 } $.
One solution is SM like with $c_3$ centers around zero. Another possible solution, $c_3 \sim -0.14$, corresponds to that the  NP gluon-gluon fusion amplitude equals minus two times the SM amplitude.
We would like to point out that this degeneracy can be easily lifted once the LHC has enough  Higgs pair production data in the high luminosity phase.
The  Higgs pair production by gluon-gluon fusion in the SM, $G^\mu(p_a)G^\nu(p_b)\rightarrow h(p_c)h(p_d)$, has been analyzed in \cite{HiggsPair-1} and summarized in the review paper \cite{Higgs-rev}.
In the SM, the two gluons must carry the same color and this process receives contributions from the triangle diagram and box diagram.
Both diagrams yield the spin-0 form factor while the box diagram contributes additional spin-2 form factor.
The SM amplitude is
\beq
\mathcal{M}(G^a G^b\to H^c H^d)= \frac{G_F\alpha_s s}{2\sqrt{2}\pi} \left [{3 m_h^2 \over s-m_h^2} F_T A_{0\mu\nu} +
(F_B A_{0\mu\nu}+
G_B A_{2\mu\nu})  \right]\epsilon^\mu_a \epsilon^\nu_b\,,
\eeq
where $s=(p_a+p_b)^2$, and the spin-0/2 form factors $A_0^{\mu\nu}(\propto S^{\mu\nu})/A_2^{\mu\nu}$ are given by
\begin{align}
&A_0^{\mu\nu}=g^{\mu\nu}-\frac{p_a^\nu p_a^\mu}{p_a\cdot p_b}\,,\\
&A_2^{\mu\nu}=g^{\mu\nu}+\frac{p_c^2 p_a^\nu p_b^\mu}{p_T^2 (p_a\cdot p_b)}
-\frac{2(p_b\cdot p_c)p_a^\nu p_c^\mu}{p_T^2 (p_a\cdot p_b)}
-\frac{2(p_a\cdot p_c)p_b^\mu p_c^\nu}{p_T^2 (p_a\cdot p_b)}+\frac{2p_c^\mu p_c^\nu}{p_T^2}\,.\nonumber
\end{align}
For more details, see \cite{HiggsPair-1}.
The operator ${\cal O}_3$ provides an additional contribution to the spin-0 amplitude, and the amplitude square becomes
\begin{equation}
|\mathcal{M}|^2=\frac{G_F^2\alpha_s^2 s^2}{8\pi^2}
\left(\left|\frac{3 m_h^2}{s-m_h^2}F_T+F_B-\frac{8\sqrt{2}\pi^2 }{G_F\Lambda^2 }c_3\right|^2
+|G_B|^2
\right)\,.
\end{equation}
If taking $\Lambda=1\mathrm{TeV}$, the exotic solution that $c_3 \sim -0.14$ gives  $-\frac{8\sqrt{2}\pi^2}{G_F\Lambda^2}c_3 \sim + 1.34$ inside the spin-0 amplitude square. Because of the cancelation between the contributions from $F_T$ and $F_B$\footnote{In the large quark mass limit $F_T=\frac23+\mathcal{O}(s/m_Q^2)$, $F_B=-\frac23+\mathcal{O}(s/m_Q^2)$,
and $G_B=\mathcal{O}(s/m_Q^2)$\cite{HiggsPair-1}.},  this exotic solution generates a sizable deviation from the SM prediction.
Therefore, this two-fold degeneracy could be resolved by the future  Higgs pair production data.

\subsection{UV complete models}
Here we discuss and make comparisons the  validity of our effective operator analysis of two UV complete toy models.
Let's first examine the model which contains an additional  color octet scalar $S$ whose SM quantum number is $(8,2,1/2)$ \cite{MW} on top of the SM.
The most general scalar potential is
\begin{eqnarray}
\label{eq:Oct_pontential}
V &=& \frac{\lambda}{4}\left(H^{\dagger i} H_i\right)^2 + 2m_S^2 \mathrm{Tr}S^{\dagger i} S_i
+\lambda_1 H^{\dagger i}H_i \mathrm{Tr} S^{\dagger j} S_j + \lambda_2 H^{\dagger i}H_j \mathrm{Tr}S^{\dagger j} S_i \nonumber\\
&& +\left(\lambda_3 H^{\dagger i}H^{\dagger j} \mathrm{Tr}S_i S_j + \lambda_4 H^{\dagger i}\mathrm{Tr}S^{\dagger j}S_j S_i
+\lambda_5 H^{\dagger i} \mathrm{Tr}S^{\dagger j}S_i S_j + h.c.\right)\nonumber\\
&& +\lambda_6 \mathrm{Tr}S^{\dagger i} S_i S^{\dagger j} S_j + \lambda_7\mathrm{Tr}S^{\dagger i} S_j S^{\dagger j} S_i
+\lambda_8\mathrm{Tr} S^{\dagger i} S_i \mathrm{Tr}S^{\dagger j} S_j\nonumber\\
&& + \lambda_9 \mathrm{Tr} S^{\dagger i} S_j\mathrm{Tr} S^{\dagger j}S_i +\lambda_{10} \mathrm{Tr}S_i S_j \mathrm{Tr}S^{\dagger i}
S^{\dagger j} +\lambda_{11} \mathrm{Tr}S_i S_j S^{\dagger j} S^{\dagger i}\,,
\end{eqnarray}
where $i,j$ are the $SU(2)$ indices, and $m_S$ is the mass of the color octet scalar.
The custodial symmetry requires the following relations to be held:
$
2\lambda_3=\lambda_2,
2\lambda_6=2\lambda_7=\lambda_{11},
\lambda_9=\lambda_{10}
$ for the real couplings\cite{MW}, and $\lambda_4=\lambda_5^*$  for the complex ones\cite{XGH-octet}.
And the SM quarks can now couple to both $H$ and $S$ by:
\begin{equation}
\label{eq:Oct_Yukawa_Gen}
-\mathcal{L}_Y=y^U_{ij}\bar{Q}_{Li}\tilde{H} u_{Rj} + y^D_{ij} \bar{Q}_{Li} H d_{Rj} + Y^U_{ij}
\bar{Q}_{Li}\tilde {S}^AT^A u_{Rj}
+  Y^D_{ij}\bar{Q}_{Li} S^A T^A d_{Rj}+ h.c.\,,
\end{equation}
where $T^A$ is the $SU(3)$ generator, and $A$ is the color index.
The Yukawa couplings $y$ and $Y$ are in general complex.
\begin{figure*}[h]
$$
\includegraphics[width=10cm]{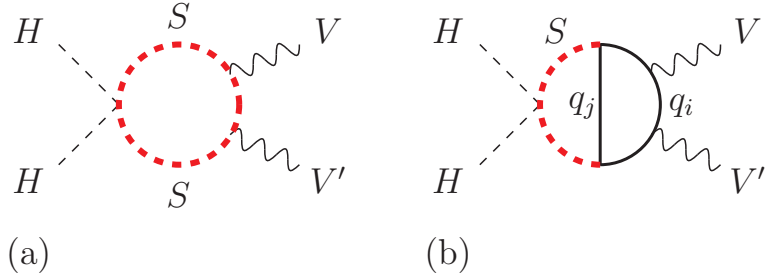}
$$
\caption{ Some typical loop diagrams which generate the effective  gauge-Higgs operators in the color octet scalar model.
Where $S$, the red(thick) dash lines, are the color octet scalars, $q$'s are the SM quarks, $H$ is the SM Higgs doublet,  $V$ and $V'$ are the SM vector bosons. Note there are many other  ways to attach the $V$ and $V'$ if gauge symmetry allowed.
Diagram (a) is the leading contribution to the CP-even gauge-Higgs operators. The CP-odd gauge-Higgs operators begin at two-loop level, diagrams (b).
  }
\label{fig:octet}
\end{figure*}
In  Fig.\ref{fig:octet} we show how we can generate
these effective gauge-Higgs operators. One typical diagram to produce CP-even operator is Fig.\ref{fig:octet}(a),
which is at one-loop level. Based on a naive dimensional analysis, the Wilson coefficients are given \cite{MW}
\begin{align}
&\frac{c_3}{\Lambda^2}=\frac32\frac{c_2}{\Lambda^2}=\frac32 \frac{c_1}{\Lambda^2}=
\frac{2\lambda_1+\lambda_2}{64\pi^2 m_S^2}\,,\nonumber\\
&\frac{c_{12}}{\Lambda^2}=\frac{\lambda_2}{48\pi^2m_S^2}\, .
\end{align}
And the CP violating operators are generated at two-loop level, see Fig.\ref{fig:octet}(b).
A ball park estimation gives
\begin{equation}
\frac{\tilde{c}_{1,2,12,3}}{\Lambda^2}\sim \sum_{i,j}\frac{\mathrm{Im}[ u^*_{ij}v_{ij}]}{(16 \pi^2)^2} \frac{ m_{q_i} m_{q_j}}{ m_S^2 \mathrm{max}\{ m^2_{q_{i,j}}, m_H^2 \} }\,.
\end{equation}
where $q_{i,j}$ are in the mass basis and  the octet Yukawa are parameterized as $\bar{q}_i T^A(u_{ij}+ v_{ij}\gamma^5)q_j S^A$.
Even though the CP violating phase is of order one, the strength of $\tilde{c}$'s are roughly two orders smaller than the CP-even Wilson coefficients in this model.
\begin{figure*}[h]
$$
\includegraphics[width=14cm]{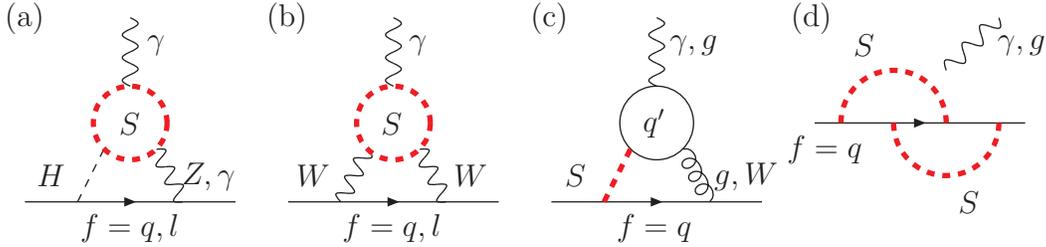}
$$
\caption{ Some typical leading loop diagrams which generate  quark or lepton AMDM, (a) and (b), and quark EDM and cEDM, (c) and (d), in the color octet scalar model.
Where  $S$, the red(thick) dash lines, are the color octet scalars, and $f$ is the SM fermion, either lepton or quark.
 Note there are many other  ways to attach the external photon if gauge symmetry allowed.
  }
\label{fig:octet_EMDM_UV}
\end{figure*}

Since this model is UV complete, we are able to discuss the fermion EDM and AMDM at above the electroweak scale without encountering any divergence.
The  charged lepton (g-2) starts at two-loop level, see Fig.\ref{fig:octet_EMDM_UV}(a,b)\footnote{ The resulting 2-loop (g-2) can be
read and translated  from \cite{2loop_g2} where  the similar diagrams due to exotic scalars have been considered.  }.
Both Fig.\ref{fig:octet_EMDM_UV}(a) and Fig.\ref{fig:octet_EMDM_UV}(b) can be related to the diagram
shown in Fig.\ref{fig:octet}(a) by substituting VEV(s) for either one or two of external $H$ legs.
And this is exactly what we have performed in the gauge-Higgs operator analysis for $\triangle a_{e}$ and $\triangle a_{\mu}$.

For the most general Yukawa coupling, Eq.(\ref{eq:Oct_Yukawa_Gen}), the SM quarks receive nonzero EDM at 2-loop level, see Fig.\ref{fig:octet_EMDM_UV}(c,d)\footnote{ The formulas in \cite{SUSY2loopEDM}, where EDMs are generated via the similar diagrams in SUSY models, can be easily translated for use in this Octet model.  }. Both Fig.\ref{fig:octet_EMDM_UV}(c) and Fig.\ref{fig:octet_EMDM_UV}(d) are independent
of Fig.\ref{fig:octet}(b), and our  gauge-Higgs operator estimations for quark EDMs are subleading.
 On the other hand, there are two kinds of contributions to the SM lepton EDM:
(1) 3-loop diagrams, by connecting Fig.\ref{fig:octet}(b) to the lepton line by either
the SM Higgs or gauge bosons. These leading contributions have been taken care of by our gauge-Higgs operator analysis.
(2) 4-loop diagrams, first by joining the two external quark lines to form a loop in Fig.\ref{fig:octet_EMDM_UV}(c) or \ref{fig:octet_EMDM_UV}(d),
and then connecting the resulting bulb to the lepton lines via either SM Higgs or gauge bosons.
 So our gauge-Higgs operator estimation for lepton EDM are indeed the leading contribution.
Nevertheless,  in the case that  Yukawa couplings are most general, one needs to incorporate the Fermion EDM/ AMDM effective operators
in the analysis, see \cite{Chang:2005wu} for an earlier study on the interplay of the 4-fermi operators and the electric  and magnetic diploe operators.

To avoid the FCNC and other phenomenological problems, it is a common practice to assume that $Y_{ij}^{U/D} = \beta^{U/D} y_{ij}^{U/D}$, where  $\beta^{U/D}$ are real. In that case,
both Yukawa are real in the fermion mass basis, and the contributions from Fig.\ref{fig:octet_EMDM_UV}(c) and Fig.\ref{fig:octet_EMDM_UV}(d) vanish. However, the CP violating gauge-Higgs operators, Fig.\ref{fig:octet}(b), vanish as well and we do not have any say about the CP violation constraint.
Assuming that $\Lambda=m_S=1\mathrm{TeV}$, our numerical gives
 $-3 \lesssim c_1 \lesssim 4$,
 $-2.5 \lesssim c_2 \lesssim 3$,  $-0.25 \lesssim c_{12} \lesssim 0.25$ (from S parameter ),  and
  $-0.02 \lesssim c_3 \lesssim 0.015$, if we take the SM-like
 solution, then
\begin{align}
-118 \lesssim \lambda_2 \lesssim 118\,, \nonumber\\
-12.6 \lesssim 2\lambda_1+\lambda_2 \lesssim 9.5 \,.
\end{align}
In addition to the positivity conditions for the scalar potential, Eq.(\ref{eq:Oct_pontential}), the above limits provide further nontrivial constrains on the parameter space of this model.

Based on the above discussion, one can tailor a modified UV complete model  where our gauge-Higgs operator analysis is applicable.
In one of the modified versions, the octet scalar is  replaced by a real $SU(2)$ triplet, $S_t$, which carries the SM
quantum number $(8,3,0)$. Because of the $SU(2)$ representation, it has no Yukawa couplings to all SM fermions
and thus there is  neither  FCNC problems nor the EDM contributions from  Fig.\ref{fig:octet_EMDM_UV}(c) and Fig.\ref{fig:octet_EMDM_UV}(d).
The most  general renormalizable potential of the Octet-triplet scalar is
\begin{eqnarray}
	V(S_t)	&=&	\frac{\lambda}{4}(H^{\dagger}H)^{2}
				+ m_{S_t}^{2}\,\text{Tr}(S_t^{i}S_t^{i})
				+i \,\mu_{t}\,\epsilon^{ijk}\,\text{Tr}(S_t^{i}S_t^{j}S_t^{k})\nonumber\\
			&&	+\lambda_{t1}H^{\dagger}H\,\text{Tr}(S_t^{i}S_t^{i})
				+\lambda_{t2}\,\text{Tr}(S_t^{i}S_t^{i}S_t^{j}S_t^{j})
				+\lambda_{t3}\,\text{Tr}(S_t^{i}S_t^{j}S_t^{i}S_t^{j})\nonumber\\
			&&	+\lambda_{t4}\,\text{Tr}(S_t^{i}S_t^{i})\,\text{Tr}(S_t^{j}S_t^{j})
				+\lambda_{t5}\,\text{Tr}(S_t^{i}S_t^{j})\,\text{Tr}(S_t^{i}S_t^{j})
\end{eqnarray}
where $i,j,k= 1,2,3$ are the $SU(2)$ indices for adjoint representation.
Moreover, we introduce a pair of
exotic vector fermions, $\psi_{L,R}$, whose SM quantum numbers are $(6,2,1/2)$.
The vector fermions admit a Dirac mass and a Yukawa coupling to $S_t$,
\beq
{\cal L}  \supset  m_\psi \bar{\psi}_L\psi_R + \eta_t \bar{\psi}_L S_t \psi_R +h.c.
\eeq
The Dirac mass $m_\psi$ is taken to be real without losing any generality and assumed to be much larger than $v$ so
that the exotic fermions decouple as $S_t$ does when energy is below the electroweak scale. The Yukawa $\eta_t$ is complex in general.
The CP violating gauge-Higgs operators  can be generated at  two-loop levels, by replacing $(S,q)\rightarrow (S_t,\psi)$ in Fig.\ref{fig:octet}(b),
and a rough estimation for the Wilson coefficients are
\begin{equation}
\frac{\tilde{c}_{1,2,12,3}}{\Lambda^2}\sim \frac{1}{(16 \pi^2)^2} \frac{ \mathrm{Re}\,\eta_t\, \mathrm{Im}\,\eta_t }{ \mathrm{max}\{m_{S_t}^2,m_\psi^2\}}\,.
\end{equation}
For $m_{S_t} \sim m_{\psi}\sim 1$ TeV,  the $\theta_{QCD}$ gives the strongest constraint that $|\eta_t|^2 |\sin 2\theta_t |<1.01\times 10^{-6}$.
The CP-even Wilson coefficients are generated at 1-loop level ( by replacing $S$ by $S_t$ in Fig.\ref{fig:octet}(a) )
\beq
\frac{4}{3} \frac{c_3}{\Lambda^2}=\frac{3}{4}\frac{c_2}{\Lambda^2} = \frac{\lambda_{t1}}{16\pi^2 m_{S_t}^2}\,,
\eeq
but nonzero $c_1$ and $c_{12}$ need to be generated at higher-loop level since $S_t$ carries no hypercharge.
And our global fit yields that $-4.2 \leq \lambda_{t1} \leq 3.2$.

\subsection{Predictions for the CP-odd decays }
\label{sec:CPodd_prediction}
The spin and parity of the 125 GeV boson has been largely determined to be $J^P=0^+$ from
the analysis of the polar angular distribution in the $h\rightarrow 2\gamma $ mode\cite{ATLAS-gamma,CMS-gamma} (for spin) and
the polarization correlation  in the $h\rightarrow ZZ\rightarrow 4l$ decay\cite{ATLAS-ZZ,CMS-ZZ} (for parity).
However, the current spin-parity analyses are all based on limited  number of events and simple working assumptions,
 namely the boson is assumed to be either purely scalar, pseudoscalar, or spin-2 object ect.
But we have to keep in mind that even the 125 GeV boson is an elementary scalar, the  CP violation decays can still
be generated by quantum corrections. For a given model, whether we have the required experimental sensitivity to detect the CP-odd
composition is the question.

If at low energy the new physics can be indeed described by the gauge-Higgs operators alone,
our numerical study predicts  interesting model-independent relations among the Higgs CP-odd decay modes.
As has been discussed in Sec.4, also from  Fig.\ref{fig:Higgs_EDM_MDM_allowed}(e,f),  the three CP-odd Wilson coefficients follow a linear relation on a line segment with end points at $\tilde{c}^2_{12}\sim 0.56/1.0$ for using the SM predictions/ LHC data as input.
For $\Lambda=1$TeV, we have (c.f.  Eq.(\ref{eq:alpha_Num}))
\beqa
& \alpha^{\not\!\! CP}_{\gamma\gamma} & \sim 3.21\, \tilde{c}^2_{12}\,, \nonr\\
& \alpha^{\not\!\! CP}_{\gamma Z} & \sim 1.92\, \tilde{c}^2_{12}\,, \nonr\\
& \alpha^{\not\!\! CP}_{ZZ} & \sim 4.3 \times 10^{-5}\,  \tilde{c}^2_{12}\,, \nonr\\
& \alpha^{\not\!\! CP}_{WW} & \sim 1.13\times 10^{-4}\, \tilde{c}^2_{12}\,.
\eeqa
Note that both  $h \rightarrow WW$ and $h \rightarrow ZZ$ channels are not sensitive to the presence of the CP-odd gauge-Higgs operators.

\begin{figure*}[htb]
$$
\begin{array}{cc}
\includegraphics[width=7cm]{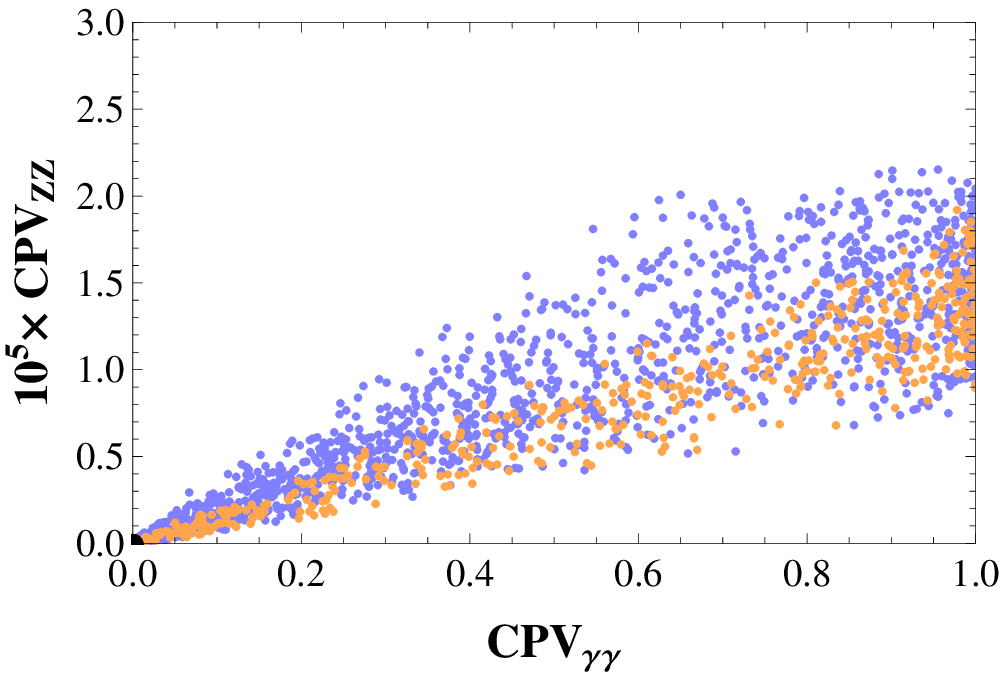}&
\includegraphics[width=7cm]{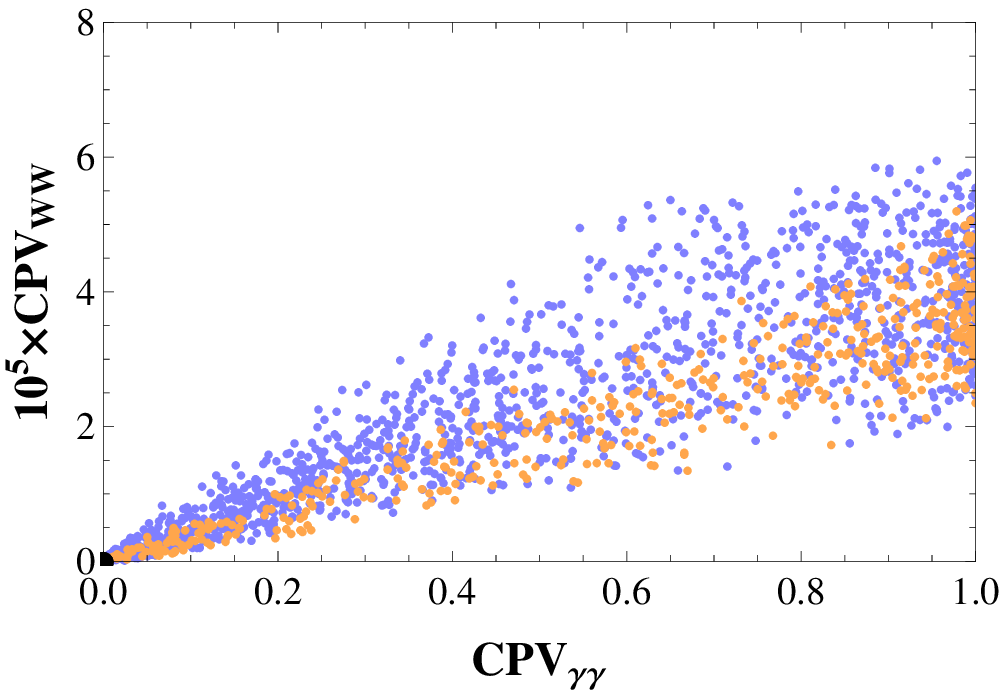}\\
(a) & (b)\\
\end{array}
$$
\caption{ The $95\%$CL CP violation fraction correlations between: (a) $h\to ZZ$ and $h \to \gamma\gamma$, and (b) $h\to WW$ and $h \to \gamma\gamma$.   The constraints are from EDMs, AMDMs, and LHC data (blue, darker) or the SM predictions (brown, lighter). }
\label{fig:CPV_aa_WW_ZZ}
\end{figure*}

We are interested in the CP violation fraction,
$ CPV_{ij} = \Gamma^{ij}_{\not\!\! CP} / \Gamma^{ij} $, which is more relevant for the parity determination.
In Fig.\ref{fig:CPV_aa_WW_ZZ}, we show the $95\%$CL CP violation fraction correlations between: (a) $h\to ZZ$ and $h \to \gamma\gamma$, and (b) $h\to WW$ and $h \to \gamma\gamma$.  Note the possible large CPV fractions due to
the gauge-Higgs effective operators in the $h \to \gamma\gamma$ and $h \to \gamma Z$ modes, which
are not constrained at all by the global fit.

Apparently,  it is not likely to  probe the CP violation fractions, at the level of $\lesssim$ ( a few )$ \times 10^{-5}$, in the $h \rightarrow WW, ZZ$ channels at the LHC\footnote{See \cite{spin-parity} for other proposals of measuring the CP nature of Higgs at the LHC.}.
This agrees with the current parity determination results\cite{ATLAS-ZZ,CMS-ZZ}. And hypothetically, any future observation of $\sim$ a few \% level CP violation fraction in the $h \rightarrow ZZ$ mode will definitely indicate the existence of  new physics beyond the gauge-Higgs sector and additional effective operators must be included.

\begin{figure*}[htb]
$$
\begin{array}{cc}
\includegraphics[width=7cm]{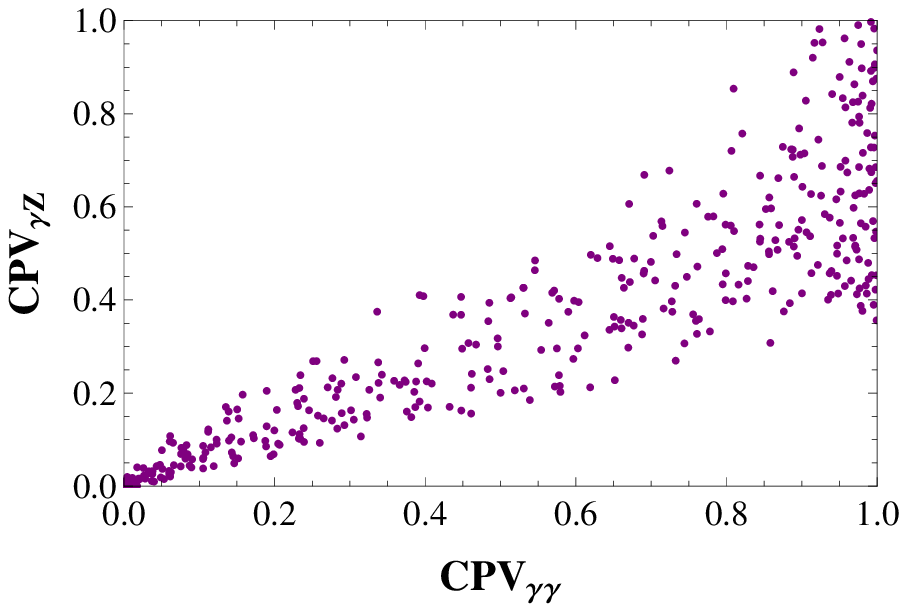}&
\includegraphics[width=7cm]{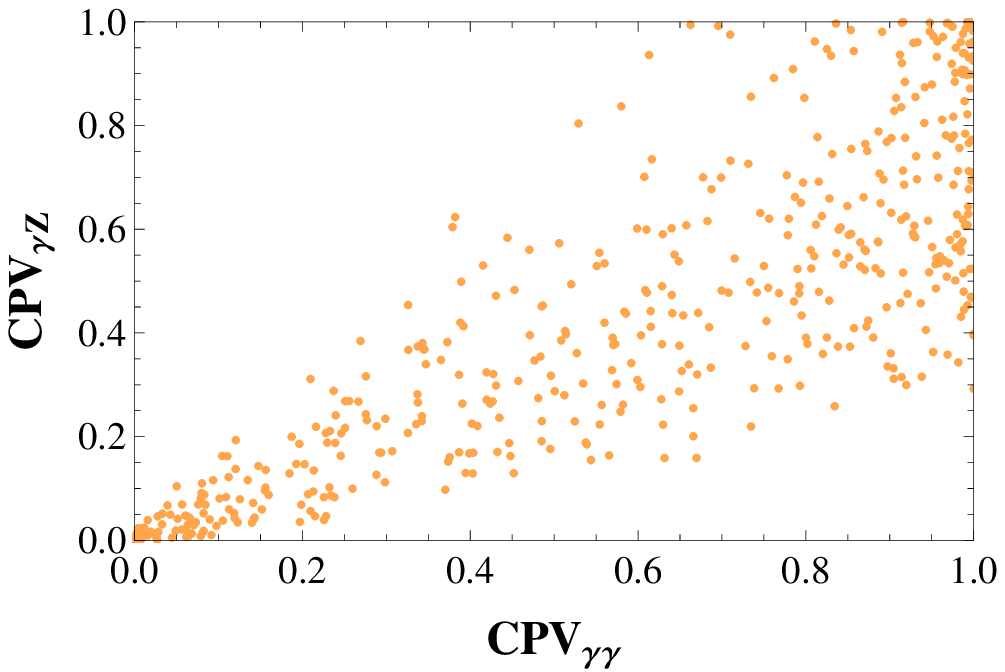}\\
(a) & (b) \\
\end{array}
$$
\caption{ (a) The $68\%$CL  and (b) the $95\%$CL  CP violation fraction correlations between $h\to \gamma\gamma$ and $h \to \gamma Z$.  The constraints are from EDMs, AMDMs, and the SM predictions. }
\label{fig:CPV_aa_aZ}
\end{figure*}

For making use of $h\rightarrow 2\gamma, \gamma Z$ channels to probe the  CP-odd components of the 125 GeV boson, we need
to study the vector boson spin correlation.  This could be   neatly done in the future $e^+e^-$, $e \gamma$, or $\gamma\gamma$ colliders, see for example \cite{photoncollider}.
For that we have a  model-independent prediction:
$\alpha_{\not\!\! CP}^{\gamma\gamma} : \alpha_{\not\!\! CP}^{\gamma Z}  \sim 1.67 :1 $.
However,  as shown in Fig.\ref{fig:CPV_aa_aZ},  when the above relation is converted into the CPV fraction, the relation is smeared
but more or less follows a linear relation.
From the plot, the CPV fraction correlation between these two modes is roughly
\beq
CPV_{\gamma z} \sim (0.65 \pm 0.35) CPV_{\gamma\gamma}\;\; (68\%\mbox{CL}).
\eeq
However, the slope becomes $(0.92 \pm 0.7)$ for the $95\%$CL fit.
Although not being very predictive, this prediction can be checked once the experimental sensitivities meet the SM theoretical uncertainties and the CPV fractions in $\gamma\gamma$ and $\gamma Z$ modes are measured in the future.

Nevertheless, from our numerical study, we found  potentially large CP violating compositions in the $h\to \gamma\gamma, \gamma Z$ decays.

\subsection{Summary}
We have studied the new physics associated with the 125 GeV boson where the NP is assumed to be characterized by the effective gauge-Higgs operators. By global fitting we studied the correlations among various Higgs to di-boson decay modes.
In addition to the updated Higgs measurements accumulated at the LHC, we also considered the case when the experimental sensitivities
are compatible with the SM theoretic uncertainties.
We found that there is  plenty of room for new physics to hide in the shadow of the inherent SM theoretical uncertainties.
From the global fit, we found a robust prediction that $\mu_{ZZ}\simeq \mu_{WW}$ and $ 0.6 \lesssim \mu_{ZZ,WW} \lesssim 1.4$ at $95\%$CL, see Figs.\ref{fig:alpha_Mu_corl_Higgs} and \ref{fig:alpha_correlation}. This can be tested at the LHC with more data in the near future.

 Moreover, we take into account the constraints of EDMs and AMDMs under the
assumption that the gauge-Higgs operators give the dominate contributions to them.
We found very strong constraints on the CP-odd Wilson coefficients and
our numerical indicated that the CP violating $hVV$ interactions
in the $h \rightarrow WW$ and  $h \rightarrow ZZ$ modes are too small, $\sim {\cal O}(10^{-5})$ at most, to be detected at the LHC, see Fig.\ref{fig:CPV_aa_WW_ZZ}.
However, the CP-odd fraction in the $h\gamma\gamma$ and $h\gamma Z$ channels could be as large as ${\cal O}(1)$.

Two more intriguing relations are predicted if the EDM constraints are included:
(1) $(\alpha_{WW}-1) \sim 2(\alpha_{ZZ}-1)$, and
(2) $(\alpha_{\gamma Z}-1)\sim 6500 (\alpha_{WW}-1)^2$.
These predictions could  be tested at  the future precision Higgs machines.

The caveats: we should keep in mind that our results are valid only for the class of NP which manifests itself in the form of the gauge-Higgs operators discussed in this work. So our results are not applicable if (1)  existence of any light exotic degree of freedom below the electroweak scale which interacts with the Higgs boson, or (2) the NP is beyond the realm which can be largely charted by the gauge-Higgs operators.
On the other hand, any violation of our above predictions will indicate that the new physics must go beyond the gauge-Higgs sector.

\begin{acknowledgments}
Work was supported by the Taiwan NSC, Grant No.
99-2112-M-007-006-MY3. FX is also supported by the Taiwan NSC, Grant No.
101-2811-M-007-051 .
\end{acknowledgments}

\appendix
\section{Decay width of $h\rightarrow  W W^*, Z Z^*$}
Details of calculating the width of $h\rightarrow V V^*$ ( $V=W, Z$) are collected in this section.
We consider the case that the virtute gauge boson $V^*$  subsequently decays into two fermions $f$ and $f'$ with their momentums labeled as: $h(P) \rightarrow V(p_1)+ f(p_2)+f'(p_3)$.
In the  center of mass frame of $f$ and $f'$, the three-body decay width is given by the following phase integral
\begin{align}
&\Gamma =\frac{1}{256 \pi^3 m_h^3}\int^{(M-m_1)^2}_{(m_2+m_3)^2} ds_1
\int^{s_{2+}}_{s_{2-}}
 ds_2 \left( |\mathcal{M}_1|^2 +|\mathcal{M}_2|^2\right)\,,\\
& s_{2\pm}=m_1^2+m_3^2+\frac{1}{2s_1}\left[
(s-s_1-m_1^2)(s_1+m_3^2-m_2^2)\pm \lambda^{\frac12}(s_1,m_3^2,m_2^2)
\lambda^{\frac12}(s_1,M^2,m_1^2)
\right]\,,\nonumber\\
&\lambda(a,b,c)\equiv a^2+b^2+c^2-2ab-2ac-2bc\,,
\end{align}
where  the kinematical variables are defined as: $s=P^2$, $p_4\equiv(p_2+p_3)$,  $s_1=p_4^2$,
$s_2=(p_1+p_3)^2$, $s_3=(p_1+p_2)^2$, and $\mathcal{M}_1$ and $\mathcal{M}_2$ are the CP-even and CP-odd amplitudes respectively
The phase space integral is largely simplified when the final state fermions are massless which is a good approximation in our case.

Beside the SM contribution, the CP-even amplitude also receives the contributions from
${\cal O}_1$ and ${\cal O}_2$ and
\begin{eqnarray}
|\mathcal{M}_{1}|^2
&=&\frac{4N_c^f(g_V^2+g_A^2)}{(s_1-m_V^2)^2}\times \left[\left(m_V^2 4\sqrt{2}G_F-a_V^{'2} s_1 \right)(2 p_1\cdot p_2 p_1\cdot
p_3 -p_1^2 p_2\cdot p_3)\right.\nonumber\\
&&\left.\hspace{3.3cm} +
(m_V^2 2(\sqrt{2}G_F)^{\frac12}-a_V^{'} p_1\cdot p_4)^2 2 p_2\cdot p_3
\right]\,,
\end{eqnarray}
where   $a'_W= 2 c_2 v g_2^2/\Lambda^2$, $a'_Z=2 a_4 v g_2^2/\Lambda^2$,  $N_c^f$ the color factor of the final state fermions,
and $(g_V, g_A)= (g_2/2 c_W)(I_3-2 Q_f s_W^2, -I_3)$ for $Z^0$ and $g_V= -g_A=g_2/2\sqrt{2}$ for $W^\pm$.
And $\mathcal{M}_{2}$ is completely beyond the SM,
\begin{eqnarray}
|\mathcal{M}_{2}|^2
= 8(g_V^2+g_A^2)\left(\frac{\tilde{a}'_V}{s_1-m_V^2}\right)^2 \times \left[(p_1\cdot p_4) (p_3\cdot p_4 p_1\cdot p_2+p_2\cdot p_4 p_1\cdot p_3)\right.\nonumber\\
\left. -p_4^2 p_1\cdot p_2 p_1\cdot p_3 - p_1^2 p_2\cdot p_4 p_3\cdot p_4 \right]\,,
\end{eqnarray}
where  $ \tilde{a}'_W= 2 \tilde{c}_2 v g_2^2/\Lambda^2$, and $ \tilde{a}'_Z=2 \tilde{a}_4 v g_2^2/\Lambda^2$.
Summing up all massless final states, the corresponding  CP-even decay widths  are:
\begin{align}
&\Gamma_{1}^W=\frac{9G_F^2m_W^4m_h}{8\pi^3}R_1(m_W/m_h,a_W')\qquad (\mathrm{for}~ W)\,,
\nonumber\\
&\Gamma_{1}^Z=\frac{3G_F^2m_Z^4m_h}{32\pi^3}
\delta_Z
R_1(m_Z/m_h,a_Z')  \qquad (\mathrm{for}~ Z)\,,
\end{align}
with $\delta_Z=\left(
7-\frac{40}{3}s_W^2+\frac{160}{9}s_W^4
\right)$, the phase space factor $R_1$ will be given in below.
 The $\delta_Z$ factor is different from \cite{Higgs-rev} and differs by a factor of 3
for W case, but our results are consistent with \cite{hVV}.

For the CP-odd part, the decay widths for W and Z bosons are
\begin{align}
&\Gamma_{2}^W=\tilde{c}_2^2\frac{9m_h^3g_2^6v^2}{32\pi^3\Lambda^4}R_2(\epsilon)
=\tilde{c}^2_2\frac{9G_F^2m_W^4m_h}{8\pi^3}\cdot \frac{32 m_h^2m_W^2}{\Lambda^4}R_2(\epsilon)\,,
\nonumber\\
&\Gamma_{2}^Z=
 \tilde{a}^{2}_4\delta_Z
\frac{3G_F^2m_Z^4m_h}{32\pi^3}\cdot\frac{32m_h^2 m_Z^2 c_W^4}{\Lambda^4}
R_2(\epsilon)\,.
\end{align}

The two functions $R_1, R_2$ can be expressed in the form of integrals,
\begin{align}
&R_1(\epsilon, a_V')\equiv \epsilon^2\int^{(1-\epsilon)^2}_0 d\tilde{s}_1
\int^{\tilde{s}_{2+}}_{\tilde{s}_{2-}}d\tilde{s}_2
\frac{1}{(\tilde{s}_1-\epsilon^2)^2}\nonumber\\
&\hspace{1cm}
\left\{\left(1-\frac{a_V^{'2}\tilde{s}_1}{4\sqrt{2}G_F\epsilon^2}\right)\left(
(1-\tilde{s}_1-\tilde{s}_2)(\tilde{s}_2-\epsilon^2)\frac{1}{\epsilon^2}
-\tilde{s}_1\right)
+2\tilde{s}_1\left(1-a'_V\frac{1-\epsilon^2-\tilde{s}_1}{4(\sqrt{2}G_F)^{\frac12}\epsilon^2}\right)^2\right\}\,,\nonumber\\
&R_2(\epsilon)\equiv \int^{(1-\epsilon)^2}_0 d \tilde s_1
\int^{\tilde s_{2+}}_{\tilde s_{2-}}d \tilde s_2
\frac{\tilde{s}_1}{(\tilde s_1-\epsilon^2)^2}
\left[\frac14(1-\epsilon^2-\tilde{s}_1)^2-\frac12\epsilon^2 \tilde{s}_1
-\frac12 (\tilde{s}_2-\epsilon^2)(1-\tilde s_1 -\tilde s_2)\right]\,,\nonumber
\end{align}
with the relevant
parameters introduced as
$\epsilon=\frac{m_V}{m_h}$,
$\tilde{s}_1=\frac{s_1}{m_h^2}$, $\tilde{s}_2=\frac{s_2}{m_h^2}$,
and $\tilde{s}_{2\pm}= \frac12(1+\epsilon^2-\tilde{s}_1)\pm\frac12
\sqrt{[(1+\epsilon)^2-\tilde{s}_1][(1-\epsilon)^2-\tilde{s}_1]} $.
The analytical form for $R_{1,2}$ can be obtained:
\begin{align}
&R_1(\epsilon,a)
=-\frac{1}{72\epsilon^2}\left[A+
 B \arctan\frac{(1-\epsilon^2)\sqrt{4\epsilon^2-1}}{1-3\epsilon^2}
+C\ln\epsilon\right]\,,
\\
& \hspace{1cm} A=(1-\epsilon^2)\left[36a \epsilon^2 (-5+9\epsilon^2)+6(2-13\epsilon^2+47\epsilon^4)
+a^2(17-82\epsilon^2+89\epsilon^4)\right]\,,\nonumber\\
& \hspace{1cm} B=\frac{6}{\sqrt{4\epsilon^2-1}}\left[6(\epsilon^2-8\epsilon^4+20\epsilon^6)+12a(\epsilon^2-8\epsilon^4
+14\epsilon^6)+a^2(-1+11\epsilon^2-40\epsilon^4+54\epsilon^6)\right]\,,\nonumber\\
& \hspace{1cm} C=6\left[6(\epsilon^2-6\epsilon^4+4\epsilon^6)+12a(\epsilon^2-6\epsilon^4+2\epsilon^6)
+a^2(-1+9\epsilon^2-30\epsilon^4+6\epsilon^6)\right]\,,\nonumber\\
&R_2(\epsilon)
=\frac{(1-\epsilon^2)}{36}\left(-17+64\epsilon^2+\epsilon^4\right)+
\frac{\ln \epsilon}{6}(1+6\epsilon^4-9\epsilon^2)\nonumber\\
&\hspace{1.5cm}
+\frac{(7\epsilon^2-1)}{6}\sqrt{4\epsilon^2-1}\arctan\frac{(\epsilon^2-1)\sqrt{4\epsilon^2-1}}{3\epsilon^2-1}\,,
\end{align}
where
\begin{equation}
a\equiv \left\{\begin{array}{ll}
\frac{8m_W^2}{\Lambda^2}c_2,& \quad \mathrm{for}\; W\\
\frac{8m_W^2}{\Lambda^2}a_4,& \quad  \mathrm{for}\; Z
\end{array}
\right.\;.
\end{equation}
One can check that  when $a'_V=0$,  the SM result is recovered, and  $6 R_1= F(\epsilon)$   as defined in\cite{hVV}.

\end{document}